\documentstyle[12pt,graphicx]{article}
\def\beqar {\begin{eqnarray}}
\def\eeqar {\end{eqnarray}}
\def\beq {\begin{equation}}
\def\eeq {\end{equation}}
\def\A{{\cal A}}
\def\B{{\cal B}}
\def\C{{\cal C}}
\def\F{{\cal F}}

\def\S{{\cal S}}

\def\P{{\cal P}}

\def\N{{\cal N}}
\def\M{{\cal M}}

\def\al{\alpha}
\def\bt{\beta}
\def\del{\delta}

\def\ga{\gamma}

\def\ep{\epsilon}
\def\la{\lambda}

\def\om{\omega}
\def\Om{\Omega}
\def\th{\theta}

\def\si{\sigma}
\def\Si{\Sigma}

\def\p{\phi}
\def\d{\partial}

\def\Ad{{\dot A}}
\def\Bd{{\dot B}}

\def\bz{{\bar z}}
\def\bu{{\bar u}}

\def\hf{\frac{1}{2}}

\def\<{\langle}\def\bra{\langle}
\def\>{\rangle}\def\ket{\rangle}

\def\Tr{{\rm Tr}}

\def\Path{{\rm P}}

\def\cp{{\bf CP}}

\begin{document}

\begin{titlepage}
\null\vspace{-62pt} \pagestyle{empty}
\begin{center}
\vspace{1.0truein}

{\Large\bf Holonomies of gauge fields in twistor space 2: \\
\vspace{.36cm}
\hspace{-.6cm}
Hecke algebra, diffeomorphism, and graviton amplitudes} \\

\vspace{1.0in} {\sc Yasuhiro Abe} \\
\vskip .12in {\it Cereja Technology Co., Ltd.\\
3-1 Tsutaya-Bldg. 5F, Shimomiyabi-cho  \\
Shinjuku-ku, Tokyo 162-0822, Japan}\\
\vskip .07in {\tt abe@cereja.co.jp}\\
\vspace{1.3in}
\centerline{\large\bf Abstract}
\end{center}

\noindent
We define a theory of gravity by constructing a
gravitational holonomy operator in twistor space.
The theory is a gauge theory whose Chan-Paton factor
is given by a trace over elements of Poincar\'{e} algebra and
Iwahori-Hecke algebra.
This corresponds to a fact that,
in a spinor-momenta formalism, gravitational
theories are invariant under spacetime translations
and diffeomorphism.
The former symmetry is embedded in tangent spaces of
frame fields while the latter is realized by a braid trace.
We make a detailed analysis on the gravitational
Chan-Paton factor and show that an S-matrix
functional for graviton amplitudes can be
expressed in terms of a supersymmetric
version of the holonomy operator.
This formulation will shed a new light on studies of
quantum gravity and cosmology in four dimensions.

\end{titlepage}
\pagestyle{plain} \setcounter{page}{2} 

\section{Introduction}

In 1989, Witten showed a remarkable relation between
the Jones polynomial of knot theory and the
topological field theory \cite{Witten:1988hf}.
More concretely, it is shown that the knot invariants arise from
partition functions of three-dimensional Chern-Simons theory, or
the so-called Witten invariants.
Among many important ideas in \cite{Witten:1988hf},
there are two particular concepts to which we would like to give attention here.
Firstly, the specific choice of Chern-Simons action has been made because
it is a simple but yet nontrivial action that preserves general covariance.
We can, in principle, use any generally covariant
field theories in search of the knot invariants.
Chern-Simons theory is convenient for this purpose
since it does not contain a metric.
If one is motivated to construct a theory of gravity and to
compute physical quantities in terms of gravitons, it is, however,
inevitable to define a metric or at least a frame field.
In such a case, general covariance is achieved by
an integration over all metrics along with a
proper definition of metrics in some theory.
This would be a key concept for the construction
of gravitational theories.\footnote{
Notice that this does not mean the exclusion of Chern-Simons theory
from gravitational theories at all. In fact, it may be the case that
Chern-Simons action emerges in some fashion after integration over all metrics.
To answer this intriguing question is one of the objectives of the present paper.}

The other concept of attention is that of braid trace,
which arises from the understanding of knot polynomials.
Braid trace is a trace over braid generators but
its physical meaning in connection with a gravitational theory is
yet to be clarified.
In \cite{Witten:1988hf}, it is argued that complete information about
braid generators can be encoded by a choice of diffeomorphism on $\cp^1 = S^2$.
The $S^2$ comes from a certain geometric surgery
of a three-dimensional manifold (from $S^2 \times S^1$ to $S^3$)
where Chern-Simons theory is defined.
This suggests that, in order to have diffeomorphism invariance in some theory,
one has to sum over all possible braid structures.
The summation is expected to be realized by a braid trace.
Thus it is important to clarify the notion of braid trace and its relation
to diffeomorphism in building a gravitational theory.

Bearing in mind these two concepts, in the present paper, we shall construct
a theory of gravity in four dimensions.
Owing to a dimensional discrepancy,
these concepts may not be applicable at first glance.
But use of twistor space can remedy the problem.
For example, if we assume that a theory is given by a Chern-Simons action
(or a variant of this action) which is defined in twistor space,
we can obtain a four-dimensional theory as follows.
We first note that partition functions of Chern-Simons theory on $\M_3$ corresponds to
current correlators of a Wess-Zumino-Witten (WZW) model on $\Si = \d \M_3$,
where $\M_3$ denotes a three-dimensional manifold and $\Si = \d \M_3$
denotes its boundary.\footnote{
As well-known, this is probably the most important
mathematical concept developed in \cite{Witten:1988hf}.}
Thus, the partition functions (or the generating functionals)
of Chern-Simons theory on $\M_3$
are encoded entirely by a two-dimensional WZW model on $\Si$.
Now twistor space $\cp^3$ can be considered as a $S^2$-bundle over four-dimensional spacetime.
Identification of $\Si$ with the $S^2$ fiber of twistor space then leads to
a WZW model whose target space is the twistor space and, from this,
one can extract four-dimensional physics \'a la Penrose.

The use of twistor space is also supported by recent developments
in the so-called twistor string theory \cite{Witten:2003nn,Berkovits:2004hg}.
Twistor string theory, as part of string theory, contains
supergravity theories but, due to the nature of twistor space,
it turns out to be quite difficult to eliminate conformal invariance.
This matter is first investigated in \cite{Berkovits:2004jj}.
Some related work can also be found in \cite{Ahn:2005es,Dolan:2008gc}.
The elimination of conformal invariance from twistor string theory is proposed
in the so-called new twistor string theories \cite{AbouZeid:2006wu}.
Even in this new approach, the extraction of Einstein supergravity
(or general relativity) from twistor string theory is not yet satisfactory,
as argued in \cite{Nair:2007md, Broedel:2009ep}.
In the present paper, we shall not follow these lines of developments
but take a more practical approach.
Namely, we take advantage of the knowledge of
graviton amplitudes and construct a gravitational theory
such that it leads to correct amplitudes by a standard
field theoretic technique.

There has been much progress in computations
of both gluon amplitudes and graviton amplitudes, accompanied with
the twistor-string developments.
It is known that the graviton amplitudes
can be obtained from the gluon counterparts.
For four-dimensional theories, this was shown by
Berends, Giele and Kuijf \cite{Berends:1988zp} who utilized
the so-called Kawai-Lewellen-Tye (KLT) relation between tree amplitudes of
closed and open string theories \cite{Kawai:1985xq}.
This relation means that, as in the gluon cases,
the so-called maximally helicity violating (MHV) amplitudes for gravitons can also
be described in a remarkably succinct form if we use a spinor-momenta formalism.
Since the graviton amplitudes do, unlike the gluon ones, break conformal invariance,
one of the peculiarities of gravity lies in the nonholomorphicity in terms of
the spinor momenta.
In this context, the MHV graviton amplitudes
provide clues for the understanding of gravitational theories and
particularly of $\N = 8$ supergravity.
Analyses of the MHV graviton amplitudes along these lines can be found in
\cite{BjerrumBohr:2005jr}-\cite{Mason:2008jy}.
Extensions of these ideas to loop
calculations of graviton amplitudes have also been carried out
(see, {\it e.g.}, \cite{Bern:2005bb}-\cite{Katsaroumpas:2009iy}).
Remarkably, these calculational developments favorably support
a long-pending question of the ultraviolet finiteness of
$\N = 8$ supergravity \cite{Bern:2007hh,Bern:2007xj,Bern:2009kd}.
This is a result of great significance for the study of $\N = 8$ supergravity.
Recently, partly motivated by these results,
there are new actions proposed for
the $\N = 8$ theory \cite{Mason:2007ct,Kallosh:2007ym}.

Relationships between
$\N = 4$ super Yang-Mills theory and $\N = 8$ supergravity
in twistor space receive much attention recently.
(For the very recent developments, see, {\it e.g.},
\cite{Bianchi:2008pu}-\cite{Bern:2009xq} and
for relatively earlier investigations,
see also \cite{Bern:1998sv}-\cite{Ananth:2007zy}.)
As mentioned above, the two theories have a structural similarity but they
also have a mathematical difference in terms of a conformal property.
Clarification of the similarity and difference will
lead to a unified way of understanding the two theories in four dimensions.
One way of having a unified picture is to regard a gravitational theory
as a gauge theory and to introduce a notion of Chan-Paton factor in the former.
This picture is in consistent with Berkovits' open-string description of twistor string
theory \cite{Berkovits:2004hg} and is first
suggested by Nair in order to interpret a physical structure of
the MHV graviton amplitudes \cite{Nair:2005iv}.
Generalization of Nair's interpretation to non-MHV amplitudes is carried out
in \cite{Abe:2005se}.
One of the advantages of this approach is that
we can encode the breaking of conformal invariance entirely in a Chan-Paton factor.

As we have postulated in an accompanying paper \cite{Abe:hol01},
any physical observables of gauge theories in twistor space
can universally be generated by a ``holonomy operator'' in twistor space.
In the present paper, we shall show that this idea also holds for a gravitational theory.
If we make use of a spinor-momenta formalism in twistor space,
Lorentz invariance is manifest, and hence,
in considering a certain representation of Poincar\'e algebra in this framework,
the representation is essentially given by translational generators.
This is in consistent with the fact that the Chan-Paton factor of
graviton amplitudes is described by combinations of the translational generators.
Diffeomorphism invariance suggests that
these generators should be furnished with generators
of braid groups or Hecke-algebra-valued quantities.
(For mathematical backgrounds of Hecke algebra, or Iwahori-Hecke algebra,
one may refer to \cite{Jones:1987dy}-\cite{Kohno:2002bk}.)
A main objective of the present paper is to show that
a Chan-Paton factor of a gravitational
holonomy operator, which can be represented by a trace over
Pincar\'e algebra and Iwahori-Hecke algebra, naturally
leads to the Chan-Paton factor of graviton amplitudes.
As in the case of gluon amplitudes, this allows us to
express an S-matrix functional of graviton amplitudes
in terms of the gravitational holonomy operator in twistor space.

The organization of this paper is as follows. In the next section,
we recapitulate the results of the accompanying paper \cite{Abe:hol01}.
We review the definition of the above mentioned holonomy operator in twistor space.
We see that Iwahori-Hecke algebra
naturally arises from the construction of the holonomy operator.
In section 3, we consider realization of diffeomorphism in the spinor-momenta formalism
and discuss that diffeomorphism invariance can be represented by a braid trace.
We also give an appropriate definition of metrics,
following Nair's interpretation of gravity as a gauge theory.
In section 4, we construct and compute a gravitational holonomy operator in twistor space.
A Chan-Paton factor of the holonomy operator is basically composed of two ingredients.
One is a sum over all possible metrics
and the other is a braid trace.
We make a detailed analysis on this Chan-Paton factor and
see that it has one-to-one correspondence with
a certain combinatoric factor in graviton amplitudes.
In section 5, utilizing the results of the previous sections, we give
an explicit expression for an S-matrix functional of graviton amplitudes.
As in the Yang-Mills case,
the S-matrix functional is expressed in terms of a supersymmetric
version of the gravitational holonomy operator.
Lastly, we shall present some concluding remarks.

\section{Review of holonomy formalism in twistor space}

In this section, we review the construction of holonomy operators in
twistor space, which has been developed in \cite{Abe:hol01}.
We shall also discuss the emergence of Iwahori-Hekcke algebra in this
formulation, following Kohno's textbook \cite{Kohno:2002bk}.

\noindent
\underline{Spinor momenta}

A holonomy operator in twistor space is defined by use
of a spinor-momenta formalism.
Spinor momenta of massless particles, such as gluons and
gravitons, are generally given by two-component complex spinors.
In terms of four-momentum $p_\mu$ $(\mu = 0,1,2,3)$, which
obey the on-shell condition
$p^2 = \eta^{\mu\nu} p_\mu p_\nu =
p_{0}^{2} - p_{1}^{2} - p_{2}^{2} - p_{3}^{2} =0$
($\eta^{\mu\nu}$ denoting the Minkowski metric ),
the spinor momenta can be expressed as
\beq
u^A = {1 \over \sqrt{p_0 - p_3}} \left(
        \begin{array}{c}
          {p_1 - i p_2} \\
          {p_0 - p_3} \\
        \end{array}
      \right) \, , ~~
\bu_\Ad  = {1 \over \sqrt{p_0 - p_3}}
    \left(
         \begin{array}{c}
           {p_1 + i p_2 } \\
           {p_0 - p_3} \\
         \end{array}
    \right)
\label{2-1}
\eeq
where both $A$ and $\Ad$ take values of $(1,2)$.
With these, the four-momentum $p_\mu$ can be parametrized as
a $(2 \times 2)$-matrix,
$p^{A}_{\, \Ad}  = (\si^\mu)^{A}_{\, \Ad} \, p_\mu \equiv u^A \bu_\Ad$
with $\si^\mu = ( {\bf 1}, \si^i)$ where $\si^i$ ($i = 1,2,3$) denotes
the $(2 \times 2)$ Pauli matrices and
${\bf 1}$ is the $(2 \times 2)$ identity matrix.
Requiring that $p_\mu$ be real, we can take $\bu_\Ad$ as a
conjugate of $u^A$, {\it i.e.}, $\bu_\Ad = (u^A)^*$.

From the above parametrization of $p_\mu$, we see that $p_\mu$ is invariant under
\beq
u^A \rightarrow e^{i \phi} u^A \, , ~~~~~ \bu_\Ad \rightarrow e^{-i \phi} \bu_\Ad
\label{2-2}
\eeq
where $\phi$ represent a $U(1)$ phase parameter.
Thus there is a phase ambiguity in the definition of $u^A$ and $\bu_\Ad$.

Lorentz transformations of $u^A$ are given by
\beq
u^A \rightarrow (g u)^A
\label{2-3}
\eeq
where $g \in SL(2, {\bf C})$
is a $(2 \times 2)$-matrix representation of $SL(2,{\bf C})$;
the complex conjugate of this relation leads to Lorentz transformations of $\bu_\Ad$.
Four-dimensional Lorentz transformations are realized by a
combination of these, that is, the four-dimensional Lorentz symmetry is
given by $SL(2,{\bf C}) \times SL(2,{\bf C})$.
Scalar products of $u^A$'s or $\bu_\Ad$'s, which are invariant under the
corresponding $SL(2,{\bf C})$, are expressed as
\beq
 u_i \cdot u_j \equiv (u_i u_j) =   \ep_{AB} u_{i}^{A}u_{j}^{B} \, , ~~~~~
 \bu_i \cdot \bu_j \equiv [\bu_i \bu_j]  = \ep^{\Ad \Bd} \bu_{i \, \Ad}
 \bu_{j \, \Bd}
\label{2-4}
\eeq
where $\ep_{AB}$ is the rank-2 Levi-Civita tensor.
This can be used to raise or lower the indices, {\it e.g.}, $u_B = \ep_{AB}u^A$.
Notice that these products are zero when $i$ and $j$ are identical.
In what follows, we can assume $1 \le i < j \le n$ without loss of generality.

For a theory with conformal invariance, such as a theory of electromagnetism
or $\N = 4$ super Yang-Mills theory, we can impose scale invariance
on the spinor momentum, {\it i.e.},
\beq
u^A \sim \la u^A \, , ~~~~~ \la \in {\bf C} - \{ 0 \}
\label{2-5}
\eeq
where $\la$ is non-zero complex number.
With this identification, we can regard
the spinor momentum $u^A$ as a homogeneous coordinate
of the complex projective space $\cp^1$.

\noindent
\underline{Twistor space}

Twistor space is defined by a four-component spinor
$Z_I =( \pi^A, v_\Ad)$ $(I = 1,2,3,4)$ where $\pi_A$ and $v_\Ad$ are two-component
complex spinors.
From this definition, it is easily understood that
twistor space is represented by the complex homogeneous coordinates of
$\cp^3$. Thus, $Z_I$ correspond to homogeneous coordinates of $\cp^3$ and satisfy
the following relation.
\beq
Z_I  \sim \la Z_I \, , ~~~~~ \la \in {\bf C} - \{ 0 \}
\label{2-6}
\eeq
In twistor space, the relation between $\pi^A$ and $v_\Ad$ is defined as
$v_\Ad = x_{\Ad A} \pi^A$.
With this relation, the condition (\ref{2-6}) is realized by the scale invariance
of $\pi^A$, as shown in (\ref{2-5}) for $u^A$.
$x_{\Ad A}$ are defined as the local coordinates on $S^4$.
This can be understood from the fact that $\cp^3$ is a $\cp^1$-bundle over $S^4$.
We consider that the $S^4$ describes a four-dimensional compact spacetime.
A flat spacetime may be obtained by considering a neighborhood of this $S^4$.
Notice that in twistor space the spacetime coordinates $x_{\Ad A}$
are emergent quantities.
Four-dimensional diffeomorphisms, {\it i.e.},
general coordinate transformations, is therefore realized by
\beq
u^A \rightarrow u'^A
\label{2-7}
\eeq
rather than $x_{\Ad A} \rightarrow {x'}_{\Ad A}$.

{\it
What is essential in the spinor-momenta formalism in twistor space is to
identify a $\cp^1$ fiber of twistor space with a $\cp^1$ on which the spinor momenta
are defined. In other words, we identify $\pi_A$ with the spinor momenta $u_A$
so that we can essentially describe four-dimensional physics
in terms of the coordinates of $\cp^1$.
}

In the spinor-momenta formalism,
the twistor-space condition $v_\Ad = x_{\Ad A} \pi^A$ is then expressed as
\beq
v_\Ad \, = \, x_{\Ad A} u^A
\label{2-8}
\eeq

A helicity of a massless particle is generally determined by the
so-called Pauli-Lubanski spin vector. In the spinor-momenta formalism,
we can also define an analog of this spin vector, which can be used
to define a helicity operator of massless particles as
\beq
h = 1 -  \hf u^A \frac{\d}{\d u^A}
\label{2-9}
\eeq
This shows that the helicity of the particle is
essentially given by the degree of homogeneity in $u$.

\noindent
\underline{Emergence of Iwahori-Hecke algebra}

We now consider the description of gluons in particular.
The Hilbert space of the spinor-momenta formalism
for the description of gluons is given by
$V^{\otimes n} = V_1 \otimes V_2 \otimes \cdots \otimes V_n$
where $V_ i$ $(i=1,2,\cdots,n)$ denotes a Fock space that creation operators
of the $i$-th particle with helicity $\pm$ act on.
Such operators can be expressed as $a_{i}^{(\pm)}$, with $(\pm)$
denoting helicities of the gluons.
The index $i$ is called the numbering index in what follows.
Notice that $a_{i}^{(-)}$ can be given by the conjugate of $a_{i}^{(+)}$,
$a_{i}^{(-)} = (a_{i}^{(+)})^*$, and vice versa.
These can be interpreted as ladder operators
which form a part of the $SL(2, {\bf C})$ algebra.
The algebra can be expressed as
\beq
[ a_{i}^{(+)}, a_{j}^{(-)}] = 2 a_{i}^{(0)} \, \del_{ij}  \, , ~~~
[ a_{i}^{(0)}, a_{j}^{(+)}] = a_{i}^{(+)} \, \del_{ij} \, , ~~~
[ a_{i}^{(0)}, a_{j}^{(-)}] = - a_{i}^{(-)} \, \del_{ij}
\label{2-10}
\eeq
where Kronecker's deltas show that
the non-zero commutators are obtained only when $i = j$.
The remaining of commutators, those expressed otherwise, all vanish.

For a system of $n$ gluons or $n$ spinor-momenta,
the physical configuration space is given by $\C = {\bf C}^n / \S_n$,
where $\S_n$ is the rank-$n$ symmetric group.
The $\S_n$ arises from the fact that gluons are bosons with
invariance under permutations of the numbering indices.
The complex number ${\bf C}$ corresponds to a local coordinate of
each spinor-momenta defined on $\cp^1$.
It is well-known that the fundamental homotopy group of $\C = {\bf C}^n / \S_n$
is given by the braid group, $\Pi_1 (\C) = \B_n$.
The braid group $\B_n$ has generators, $b_1 , b_2 , \cdots ,
b_{n-1}$.
Let $\rho ( b_{i})$ denote a linear representation of the
braid generator $b_i$. An action of $\rho ( b_{i})$ on the Hilbert space
$V^{\otimes n}$ can basically be carried out by transposition of the index
$i$ with $i+1$.

Now, mathematically, a linear representation of a braid gruop
is equivalent to a monodromy representation of the
Knizhnik-Zamolodchikov (KZ) equation.
The KZ equation is an equation that a function on $\C$ satisfies in
general. We can denote such a function as $\Psi (z_1 , z_2 , \cdots , z_n)$,
where $z_i$ represents the local coordinate corresponding to the spinor momentum $u_i$
($i = 1, 2, \cdots, n$).
The KZ equation is then expressed as
\beq
\frac{\d \Psi }{ \d z_i} =  \frac{1}{\kappa}
\sum_{~ j \, (j \ne i)} \frac{\Om_{ij} \Psi}{z_i - z_j}
\label{2-11}
\eeq
where $\kappa$ is a non-zero constant called the KZ parameter.
We now introduce logarithmic differential one-forms
\beq
\om_{ij} = d \log (z_i - z_j) = \frac{ d z_i - d z_j}{z_i - z_j} \, .
\label{2-12}
\eeq
Notice that these satisfy the identity
\beq
\om_{ij} \wedge \om_{jk} + \om_{jk} \wedge \om_{ik} + \om_{ik} \wedge \om_{ij} = 0
\label{2-13}
\eeq
where the indices are ordered as $i < j < k$.
$\Om_{ij}$ in the KZ equation is a bialgebraic operator. In terms of the
operators of $SL(2, {\bf C})$ algebra in (\ref{2-10}), this can be
defined as
\beq
\Om_{ij} = a_{i}^{(+)} \otimes a_{j}^{(-)} + a_{i}^{(-)} \otimes a_{j}^{(+)}
+ 2 a_{i}^{(0)} \otimes a_{j}^{(0)}
\label{2-14}
\eeq
Should we have $i = j$, this would become the quadratic Casimir of $SL(2, {\bf C})$ algebra
which acts on the $i$-th Fock space $V_i$.
Introducing the following one-form
\beq
\Om =  \frac{1}{\kappa} \sum_{1 \le i < j \le n} \Om_{ij} \, \om_{ij} \, ,
\label{2-15}
\eeq
we can rewrite the KZ equation (\ref{2-11}) as a differential equation
\beq
D \Psi = (d - \Om) \Psi = 0
\label{2-16}
\eeq
where $D = d - \Om$ can be regarded as a covariant exterior derivative.

From an explicit form of (\ref{2-14}), we can show the following relations.
\beqar
\label{2-17}
[ \Om_{ij} , \Om_{kl} ] &=& 0  ~~~~~ \mbox{($i,j,k,l$ are distinct)} \\  \label{2-18}
[ \Om_{ij} + \Om_{jk} , \Om_{ik} ] &=& 0  ~~~~~ \mbox{($i,j,k$ are distinct)}
\eeqar
In mathematical literature,
these relations are called infinitesimal braid relations.
Remarkably, by use of these relations along with (\ref{2-13}),
the flatness of $\Om$, {\it i.e.}, $d \Om - \Om \wedge \Om = 0$, can be shown.
(For a proof of this, see \cite{Abe:hol01,Kohno:2002bk}.)
Therefore, it is possible to define a holonomy of $\Om$, which
gives a general linear representation of a braid group on
the Hilbert space $V^{\otimes n}$. This is the monodromy representation
of the KZ equation. The Hilbert space $V^{\otimes n}$
can then be identified as the space of conformal blocks for the KZ equation.

In physics, we need to use unitary irreducible representations (UIRs)
of certain representations. In the case of the monodromy
representation, this can by given by the so-called Iwahori-Hecke algebra
\cite{Jones:1987dy,Tsuchiya:1987rv}. In terms of
elements $\tilde{b}_i$ ($i = 1, 2, \cdots , n-1$), this algebra is
defined by
\beqar
&&
\nonumber
\tilde{b}_i \tilde{b}_{i+1} \tilde{b}_i = \tilde{b}_{i+1} \tilde{b}_i \tilde{b}_{i+1}
\, ,~~~ \mbox{if}~ |i - j| = 1 \\
&&
\tilde{b}_i \tilde{b}_j = \tilde{b}_j \tilde{b}_i
\, , ~~~~~~~~~~~~~~~ \mbox{if}~ |i - j| > 1 \label{2-19}\\
&&
\nonumber
(\tilde{b}_i - q^{\frac{1}{2} } ) (\tilde{b}_i + q^{- \frac{1}{2} }) = 0
\eeqar
where $q = \exp (i 2 \pi  / \kappa )$ and $\tilde{b}_{n}$ is identified with
$\tilde{b}_1$.
The first two relations are equivalent to the relations satisfied by
the generators $b_i$ of a braid group.
Remember that we denote a linear representation of $b_i$ as $\rho (b_i )$.
We now introduce a scaled representation $\tilde{\rho}(b_i ) = \eta \rho (b_i )$, with
$\eta = q^{1/4} = \exp ( i \pi / 2 \kappa )$.
It is known that the elements of $\tilde{\rho} (b_i )$ satisfy the last relation
of (\ref{2-19}).
This can be shown by impositions of irreducibility on each of the Fock space $V_i$.
(For details of this fact, one may refer to \cite{Kohno:2002bk}.)
Further, we may naturally impose a unitary condition
$\tilde{b}_{i}^{-1} = \tilde{b}_{i}^{\dagger}$.
Thus, the Iwahori-Hecke algebra (\ref{2-19}) forms a UIR of
a linear representation of braid groups, and this algebra should be
encoded in the definition of a holonomy operator.

\noindent
\underline{Comprehensive gauge fields and integrability}

We now introduce a ``comprehensive'' gauge one-form for the description
of $n$ gluons in the spinor-momenta formalism.
We define the comprehensive gauge field operator $A$ as
\beqar
A &=&  g \sum_{1 \le i < j \le n} A_{ij} \, \om_{ij}
\label{2-20}\\
A_{ij} &=& a_{i}^{(+)} \otimes a_{j}^{(0)} + a_{i}^{(-)} \otimes a_{j}^{(0)}
\label{2-21} \\
\om_{ij} & = & d \log(u_i u_j) = \frac{d(u_i u_j)}{(u_i u_j)}
\label{2-22}
\eeqar
where $g$ is the coupling constant.
Notice that, from the explicit form of $A_{ij}$, we can also show that
the bialgebraic quantity $A_{ij}$ satisfy the relations (\ref{2-17}) and (\ref{2-18}).
(For details of this fact, see \cite{Abe:hol01}.)
These relations are the only conditions
for the flatness or integrability of $A$.
Thus, as in the case of $\Om$, we can also obtain the expression
\beq
DA = dA - A \wedge A = - A \wedge A = 0
\label{2-23}
\eeq
where $D$ is now a covariant exterior derivative $D = d - A$.
This relation guarantees the existence of holonomies for the comprehensive
gauge field $A$.

Although the bialgebraic structures of  $\Om$ and $A$ are
different, the constituents of these remain the same, {\it i.e.}, they are given
by $a_{i}^{(0)}$ and $a_{i}^{(\pm)}$. Thus, we can use the same Hilbert space $V^{\otimes n}$
and physical configuration $\C$ for both $\Om$ and $A$.
The KZ equation of $A$ is then given by $D \Psi = (d - A) \Psi = 0$, where
$\Psi$ is a function of a set of spinor momenta $(u_1 , u_2 , \cdots u_n)$.
This suggests that the coupling constant $g$ is related to the KZ parameter $\kappa$ by
\beq
g = \frac{1}{\kappa}
\label{2-24}
\eeq

\noindent
\underline{Holonomy operators}

A holonomy of $A$ can be given by a general solution to the KZ equation
$D \Psi = (d - A) \Psi = 0$.
The construction is therefore similar to that of Wilson loop operators.
In the present formalism, rank-$n$ differential manifolds
are physically relevant for the construction.
Thus, we need differential $n$-forms
in terms of $A$ in order to define an appropriate holonomy operator.
Further, an analog of Wilson loop should be defined on $\C$.
These requirements lead to the following definition of the holonomy operator.
\beq
\Theta_{R, \ga}^{(A)} (u) = \Tr_{R, \ga} \, \Path \exp \left[
\sum_{m \ge 2} \oint_{\ga} \underbrace{A \wedge A \wedge \cdots \wedge A}_{m}
\right]
\label{2-25}
\eeq
where $\ga$ represents a closed path on $\C$ along which the integral is
evaluated and $R$ denotes the representation of the gauge group.
The color degree of freedom (or the Chan-Paton factor)
can be attached to the physical operators $a_{i}^{(\pm)}$ in (\ref{2-21}) as
\beq
a_{i}^{(\pm)} = t^{c_i} \, a_{i}^{(\pm)c_i}
\label{2-26}
\eeq
where $t^{c_i}$'s are the generators of the gauge group in the $R$-representation.
Since here we are interested in the description of gluons, the
relevant gauge groups are $SU(N)$;
we shall later consider gauge groups which are relevant to gravitons.
The symbol $\Path$ denotes an ordering of the numbering indices.
The meaning of the
action of $\Path$ on the exponent of (\ref{2-25}) can explicitly be written as
\beqar
\nonumber
\Path \sum_{m \ge 2}  \oint_{\ga} \underbrace{A \wedge \cdots \wedge A}_{m}
&=& \sum_{m \ge 2} \oint_{\ga}  A_{1 2} A_{2 3} \cdots A_{m 1}
\, \om_{12} \wedge \om_{23} \wedge \cdots \wedge \om_{m 1} \\
\nonumber
&=& \sum_{m \ge 2}  \frac{1}{2^{m+1}} \sum_{(h_1, h_2, \cdots , h_m)}
(-1)^{h_1 + h_2 + \cdots + h_m} \\
&& ~~~ \times \,
a_{1}^{(h_1)} \otimes a_{2}^{(h_2)} \otimes \cdots \otimes a_{m}^{(h_m)}
\, \oint_{\ga} \om_{12} \wedge \cdots \wedge \om_{m1}
\label{2-27}
\eeqar
where we use an ordinary definition of commutators for
bialgebraic operators. In the above expression,
$h_{i} = \pm = \pm 1$ ($i=1,2,\cdots, m$) denotes
the helicity of the $i$-th particle.
From the above expression, we can easily find that
the exponent of the holonomy operator in (\ref{2-25}) vanishes
if $m \le 1$. This explains the condition $m \ge 2$ in (\ref{2-25}).

The trace $\Tr_{R, \ga}$ in the definition (\ref{2-25}) means
traces over the Lie-algebra-valued $t^{c_i}$'s and over
the Hecke-algebra-valued $\tilde{b}_i$'s.
Since the braid generators $b_i$'s essentially give
the same algebra as $\tilde{b}_i$'s, except the last equation in (\ref{2-19}),
we can think of the trace over $\tilde{b}_i$'s as a braid trace.\footnote{
As discussed before, the difference between $b_i$ and $\tilde{b}_i$
is given by the numeric factor $\eta = \exp ( i \pi / 2 \kappa)$.
This factor will be important for some particular solutions
to a theory of gravity that we aim for, however, at the level of
trace calculations, this difference seems irrelevant and
we shall not discuss its effects in the present paper.}
Information of braid generators along the loop $\ga$
can be characterized by orderings of the numbering indices.
A braid trace is therefore realized by a sum over
permutations of the indices.
Thus the braid trace $\Tr_\ga$ over the exponent of (\ref{2-25}) can be
expressed as
\beq
\Tr_{\ga} \Path \sum_{m \ge 2}^{\infty} \oint_{\ga}
\underbrace{A \wedge \cdots \wedge A}_{m}
= \sum_{m \ge 2}
\sum_{\si \in \S_{m-1}} \oint_{\ga}  A_{1 \si_2} A_{\si_2 \si_3} \cdots A_{\si_m 1}
\, \om_{1 \si_2} \wedge \om_{\si_2 \si_3} \wedge \cdots \wedge \om_{\si_m 1}
\label{2-28}
\eeq
where the summation of $\S_{m-1}$ is taken over the
permutations of the elements $\{2,3, \cdots, m \}$,
with the permutations labeled by
$\si=\left(%
\begin{array}{c}
  2 \, 3 \cdots m \\
  \si_2 \si_3 \cdots \si_m \\
\end{array}%
\right)$.
Notice that the expression (\ref{2-28}) is valid
for a single distinct loop $\ga$.
For an $SU(N)$ gauge group, the color structure, {\it i.e.}, the trace
$\Tr_{R} ( t^{c_1} t^{c_{\si_2}} t^{c_{\si_3}} \cdots t^{c_{\si_m}} )$ with
a sum over permutations, can be characterized by one index,
say $1$, due to cyclic invariance.
Alternatively, this may be seen as a $U(1)$ invariance
of the Chan-Paton factor for Yang-Mills theory.
If we consider a gravitational theory in the same framework,
a corresponding Chan-Paton factor is expected to
have a larger symmetry, which will lead to two or
more distinct loops associated with a braid trace.
We shall clarify these points in section 4.

\section{Diffeomorphism, braid trace and Chan-Paton factors}

In this section, we start constructing a gravitational theory
by use of the spinor-momenta formalism in twistor space.
A main objective of the present paper is to express
an S-matrix functional for graviton amplitudes in terms of
a gravitational version of the holonomy operator (\ref{2-25}),
which we shall discuss in the following sections.
In this section, we develop fundamental ingredients for the construction
of a gravitational theory in the spinor-formalism.
A basic idea we will follow is Nair's interpretation that
the so-called maximally helicity violating (MHV) graviton amplitudes
can be understood as amplitudes of gauge theory (or open string theory)
with an appropriate choice of a Chan-Paton factor.
What to be clarified is then Chan-Paton factors of gravitons
in the spinor-momenta formalism, which is a main theme of this section.

\noindent
\underline{Diffeomorphism}

We first notice that the holonomy operator (\ref{2-25})
is described by differential forms.
Generally, the antisymmetrization of covariant indices and the
use of exterior derivatives are required
for the invariance under general coordinate transformations (or diffeomorphism).
So the use of holonomy operator seems to be more natural in gravitational
theories than in Yang-Mills theories.
In the present formalism, a situation is not so straightforward.
Namely, the bases of the ``covariant'' differential forms are
given by the Lorentz-invariant scalar products
$(u_i u_j)$ as shown in (\ref{2-22}). Thus an ordinary prescription
for diffeormophism by use of differential forms does not necessarily apply
to the present case. Indeed, as discussed in (\ref{2-7}), the
four-dimensional diffeomorphism is given by changes of spinor momenta or
permutations of the numbering indices for spinor momenta.
Invariance under diffeormorpism is then realized by taking
a sum over the all possible permutations.
This is nothing but a braid trace we have discussed in the previous section.
Therefore, in this sense,
the use of holonomy operator (\ref{2-25}) is appropriate for
a gravitational theory as well.
The difference from Yang-Mills theory is that gauge fields for gravitons
are not given by one-forms but by two-forms
so that they are to contain states of $\pm 2$ helicities.
These fields may be constructed as a product of ``comprehensive'' frame
fields which we consider are analogs of the comprehensive gauge fields $A$.

\noindent
\underline{Comprehensive frame fields in twistor space}

Gravitons are operators corresponding to metric tensors.
The metric tensors are generally defined by products of frame fields or tetrad fields.
In a conventional field theory, this can be expressed as
\beq
g_{\mu \nu} \, = \, e_{\mu}^{a} \, e_{\nu}^{a}
\label{3-1}
\eeq
where $g_{\mu \nu}, \, e_{\mu}^{a}$ are the metric tensors and
the frame fields, respectively.
The index $\mu$ ($= 0, 1,2,3$) denotes the Minkowski indices as before
and $a$ ($=0, 1,2,3$) denotes the coordinate indices for the tangent space.
In view of gravity as a gauge theory, its gauge group is
given by the Poincar\'{e} algebra.
An ordinary covariant derivative is then expressed as
\beq
D_\mu = \d_\mu + i \, e^{a}_{\mu} \, p^a + \Om_{\mu}^{a b} J^{a b}
\label{3-2}
\eeq
where $\Om_{\mu}^{a b}$ and $J^{a b}$ are
the spin connection and the Lorentz generator, respectively.
$\d_\mu = \frac{\d}{ \d x^\mu }$ is a differential operator with respect
to the spacetime coordinate $x^\mu$, while
$i p^a = \frac{\d} {\d x_{a} }$ is a differential operator with respect to
the tangent space coordinate.
The latter can be interpreted as a Chan-Paton factor for the frame field.

In the spinor-momenta formalism, Lorentz invariance is manifest.
Thus the last term in (\ref{3-2}) is irrelevant in the calculations of
physical quantities.
Scalar fields are by definition Lorentz invariant.
Thus another interpretation is that physical fields in the spinor-momenta formalism
are described by scalar fields (or superfields) in twistor space.
As discussed in \cite{Abe:hol01}, this is true in the Yang-Mills case and
we shall use the same formalism for gravitational cases.

We now define the comprehensive frame field $E$ as an analog of the
comprehensive gauge field $A$ in (\ref{2-20}).
\beqar
E &=& \sum_{1 \le i < j \le n} E_{ij} \om_{ij}
\label{3-3}\\
E_{ij} &=& e_{i}^{(+)} \otimes e_{j}^{(0)}
+ e_{i}^{(-)} \otimes e_{j}^{(0)}
\label{3-4}\\
\om_{ij} &=& d \log (u_i u_j) = \frac{ d (u_i u_j) }{(u_i u_j) }
\label{3-5}
\eeqar
where
{\it $e_{i}^{(\pm)}$ and $e_{i}^{(0)}$
are operators which are algebraically the same as $a_{i}^{(\pm)}$ and $a_{i}^{(0)}$,
obeying the $SL(2 , {\bf C})$ algebra in $(\ref{2-10})$.}
$\om_{ij}$ is a logarithmic one-form in terms of the
Lorentz invariant product of spinor momenta $u_i$ and $u_j$.
This one-form is the same as the Yang-Mills version in (\ref{2-22}).
We now consider a Chan-Paton factor of the frame field.
Following the Yang-Mills case, we may impose this factor on
the operators $e_{i}^{(\pm)}$ as
\beq
e_{i}^{(\pm)} = e_{i}^{(\pm) a} (\sqrt{2} p_{i})^{a}
= e_{i}^{(\pm) A \Ad} \, p_{i}^{A \Ad}
\label{3-6}
\eeq
where we split the tangent-space index $a$ ($= 0,1,2,3$)
into the two-component indices $A$ and $\Ad$.
Generally, the tangent space is given by a copy of the
coordinate space. So $p_{i}^{A \Ad}$ can be
represented by the spinor momenta of interested particles.
Explicitly, this can be written as
\beq
p_{i}^{A \Ad} = (\si^{a})^{A \Ad} p_{i \, a}
= u_{i}^{A} \bu_{i}^{\Ad }
\label{3-7}
\eeq
where, as discussed before, $\si^{a}$ is given by
$\si^{a}=( {\bf 1} , {\vec \si} )$;
${\vec \si}$ and ${\bf 1}$ denote the ordinary $(2 \times 2)$ Pauli matrices and
the $(2 \times 2)$ identity matrix, respectively.
In order to define gravitons in analogy with (\ref{3-1}),
we need to consider products of the tangent-space translational operators.
We denote these products as
\beq
p_{i}^{A \Ad} p_{j \, \Ad A} = (u_i u_j) [\bu_i \bu_j] = 2 p_{i}^{a} p_{j \, a}
\equiv \bra p_i \cdot p_j \ket
\label{3-8}
\eeq
where we use the expressions in (\ref{2-4}).

So far, we have ignored an effect of a braid generator. Namely, we
have not considered a Hecke-algebra-valued quantity in the expression (\ref{3-6}).
This is natural since a braid generator, by definition, emerges only in a
multi-particle system. As long as we consider a single frame field,
effects of braid generators are hidden.
In the definition of the comprehensive frame field (\ref{3-3}), however,
we implicitly consider a multi-particle, if not multigraviton, system.
Thus these effects are expected to be perceptible.\footnote{
These effects are in fact buried in a multigluon system as well.
In the Yang-Mills case,
a multigluon Chan-Paton factor $\Tr_{R} ( t^{c_1} t^{c_{\si_2}} t^{c_{\si_3}}
\cdots t^{c_{\si_n}} )$ is invariant in its form under permutations of
$\si$'s. ($t^{c}$'s are generators of a gauge group in
the $R$ representation.) Thus this factor is not affected by braid generators.
This is a main reason why these effects are not transparent in Yang-Mills theory.}
This point may be obvious if we try to construct a graviton field
in terms of the comprehensive frame field $E$.
Following the relation (\ref{3-1}), we can naively define
a comprehensive graviton field as a {\it product} of $E$'s.
This definition by itself however leads to a rather chaotic quantity
because a particular frame field potentially couples to any
other frame fields in tangent spaces.
A Chan-Paton factor of a single graviton should
therefore be determined  by a certain rule for the couplings.
Such a rule can and should be encoded by braid generators.
In other words, an explicit form of a graviton Chan-Paton factor
depends on a permutation of the numbering indices.

\noindent
\underline{Comprehensive fields for gravitons}

From the above argument, we find that permutations of the numbering indices
are involved in an explicit descriptions of comprehensive fields for gravitons.
For the moment, we {\it assume} that such
permutations are given by
$\si=\left(%
\begin{array}{c}
  2 \, 3 \cdots r \\
  \si_2 \si_3 \cdots \si_r \\
\end{array}%
\right)$ like the Yang-Mills case.
As we will see later, a way of taking a permutation
in a gravitational theory is not as simple as in the Yang-Mills case.
In the present section, for simplicity of discussion, we first assume
the Yang-Mills type permutation.
Notice that we are going to construct a theory of gravity
in a holonomy formalism. This means that a theory, and hence a
graviton field, is not well-defined
until a gravitational holonomy operator is constructed.
Thus a full form of a comprehensive graviton field becomes transparent,
once we define a gravitational holonomy operator in the next section,
where the full form is obtained in (\ref{4-7}).
To remind us of this fact, we shall use $r$ instead of $n$ as the
number of gravitons for the rest of this section.

Using the notation of (\ref{3-8}), we can define
a comprehensive graviton field as
\beqar
H &=& \sqrt{8 \pi G_{N}} \, \bra E \cdot E \ket
~=~ \sqrt{8 \pi G_{N}} \, \sum_{1 \le i < j \le r} H_{ij} \, \om_{ij}
\label{3-9} \\
\nonumber
\!\! H_{ij} &=& \!\!
\sum_{\si \in \S_{r-1}}  \bra E_{ij} \cdot  E_{\si_i \si_j} \ket
\, \om_{\si_{i+1} \si_{j+1}} \\
&=& \!\!
\left[ e_{i}^{(+) a} \otimes e_{j}^{(0)} +  e_{i}^{(-) a} \otimes e_{j}^{(0)} \right]
\! \sum_{\si \in \S_{r-1}}  T^{\si_i}
\left[ e_{\si_i}^{(+) a} \otimes e_{\si_j}^{(0)} +
e_{\si_i}^{(-) a} \otimes e_{\si_j}^{(0)} \right]  \om_{\si_{i+1} \si_{j+1}}
\label{3-10}
\eeqar
where $G_{N}$ is the Newton constant.
In the natural unit ($ c = \hbar = 1$), this is equivalent to
an inverse square of the Planck mass $M_{Pl}$.
\beq
G_{N} = \frac{1}{M_{Pl}^{2}} = 6.7088 \times 10^{-39} \,
\left[ \frac{1}{\mbox{GeV}} \right]^2
\label{3-11}
\eeq
In (\ref{3-10}), $T^{\si_i}$ $(i = 2, 3, \cdots , r)$ are defined as
\beqar
T^{\si_{i}} \! &=& \!\! \left\bra (p_1 + p_{\si_{i+1 < i}} + p_{\si_{i+2 < i}} +
\cdots + p_{\si_{r < i}} ) \cdot p_{\si_i} \right\ket
\, = \,
\left\langle \left( p_1 + \sum_{k=i+1}^{r} p_{\si_{k < i}} \right)
\cdot p_{\si_{i}} \right\rangle
\label{3-12}\\
p_{\si_{i<j}} & \equiv &  \left\{
    \begin{array}{ll}
      p_{\si_i} & \mbox{for $\si_i < \si_j$}\\
      0  & \mbox{otherwise}
    \end{array} \right.
\label{3-13}
\eeqar
In the case of $i= 1$, we set $\si_{1} = \si_{r+1} = 1$ and,
for a reason we discuss soon, we can define $T^{\si_{1}}$ as
\beq
T^{\si_{1}} \, = \, T^{1} \, = \, 1
\label{3-14}
\eeq

\begin{figure} [htbp]
\begin{center}
\includegraphics[width=110mm]{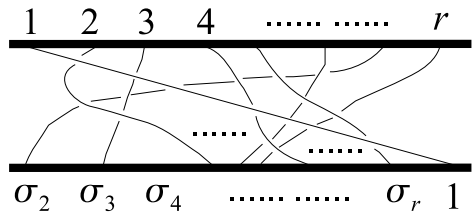}
\caption{Braid diagram associated with a permutation of
$\{ \si_2 , \si_3 , \cdots , \si_r \}$}.
\label{fig}
\end{center}
\end{figure}

{\it What is important in the definition of
$T^{\si_i}$'s is nothing but one-to-one correspondence between
$T^{\si_i}$'s and the permutations $\si$, which define how
a Chan-Paton factor of a particular frame field couples with Chan-Paton factors
of the other frame fields.}
(As we discuss at the end of section 4, the definition of
$T^{\si_i}$'s is inspired by an explicit form of graviton amplitudes.)
A pattern of a particular permutation can diagramatically be
shown as Figure \ref{fig}.

The strands in Figure \ref{fig} connect the same numbering elements
at the top and the bottom.
There is correspondence between this two-dimensional diagram
and the permutation $\si$.
We call this diagram a braid diagram in what follows.
Of course, there are many ways of drawing the strands
with arbitrary twists and turns but it is possible
to have an irreducible representation of the diagram for each permutation.
In fact, as we shall see in the next section,
the definition (\ref{3-12}) corresponds to such an
irreducible diagram under a condition that
$T^{\si_i}$ are non-vanishing.
That the numbering elements $1$ are diagonally located in
Figure \ref{fig} is due to this condition, which is also
reflected in the appearance of the term $\bra p_1 \cdot p_{\si_i} \ket$
for each of $T^{\si_i}$'s in (\ref{3-12}).

There is arbitrariness in the definition of
(\ref{3-10}) particularly in the choice of
the factor $\om_{\si_{i+1} \si_{j+1}}$.
Notice that a graviton operator is not well-defined
until a holonomy operator of $H$ is constructed.
Once a gravitational holonomy operator is
defined, the arbitrariness in (\ref{3-10}) is resolved.
The particular choice of $\om_{\si_{i+1} \si_{j+1}}$
is then justified within a holonomy formalism in twistor space;
we shall present precise definition and computation of
a gravitational holonomy operator in the next section.

As in the Yang-Mills case, a comprehensive frame field $E$ is defined
in the configuration space $\C = {\bf C}^{n}/ \S_n$.
Thus the physical configuration space for $H \sim \bra E \cdot E \ket$
is given by $\C \times \C$.
Correspondingly, $H_{ij}$ can be interpreted as two copies of bialgebraic
operators, rather than 4-algebra, each copy acting on
a distinct Hilbert space $V^{\otimes n}$.
In the operator level, gravitons can be represented by
\beq
g_{i \si_i}^{(\pm \pm)} \, \equiv \, e_{i}^{(\pm) a} \, e_{\si_i}^{(\pm) a}
\label{3-15}
\eeq
where a composite notation $(\pm \pm)$ takes any pairs.
Namely, we have $g_{i \si_i}^{(++)}$,
$g_{i \si_i}^{(+-)}$, $g_{i \si_i}^{(-+)}$ and $g_{i \si_i}^{(--)}$
among which the first and the last are relevant to gravitons
with $\pm 2$ helicities.
By use of these, we can express (\ref{3-10}) as
\beqar
\nonumber
H_{ij} &=& E_{ij} \sum_{\si \in \S_{r-1}}  T^{\si_i} \,
E_{\si_i \si_j} \, \om_{\si_i \si_j} \\
&=& \sum_{\si \in \S_{r-1}}  T^{\si_i}
\, \left[ g_{i \si_i}^{(++)} +  g_{i \si_i}^{(+-)} +
g_{i \si_i}^{(-+)} + g_{i \si_i}^{(--)}
 \right]
\, \otimes e_{j}^{(0)} \,  e_{\si_j}^{(0)} \, \om_{\si_{i+1} \si_{j+1}}
\label{3-16}
\eeqar
where we should note that
$e_{j}^{(0)}$ and $e_{\si_j}^{(0)}$
act on $e_{i}^{(\pm)}$ and $e_{\si_i}^{(\pm)}$ from the left, respectively.

We now consider the exceptional case in which $i$ becomes $i = 1$.
In the Yang-Mills case, a multigluon Chan-Paton factor
is given by $\Tr_{R} ( t^{c_1} t^{c_{\si_2}} t^{c_{\si_3}}
\cdots t^{c_{\si_n}} )$ where $t^{c}$'s are generators of a gauge group in
the $R$ representation.
In this case, permutations are taken over
$t^{c_{\si_i}}$'s $(i = 2, 3, \cdots , n$) and we can interpret
$t^{c_1}$ as a $U(1)$ direction attached to the gauge group $SU(N)$
or a $U(1)$ generator of the gauge group $U(N)$.
Thus $t^{c_1}$ can be expressed as the identity matrix
in the $R$ representation of the $U(N)$ group.
There are no braid generators associated with $t^{c_1}$.
This is consistent with the fact that the number of the elements
of a braid group $\B_n$ is given by $n-1$ rather than $n$.
We can make an analogous argument for a gravitational case.
There are no braid generators associated with the graviton labeled by the index $1$.
Thus the Chan-Paton factor of this graviton
can analogously be interpreted as an identity.
This explains the definition in (\ref{3-14}).
In terms of $H_{1j}$, this can explicitly be written as
\beq
H_{1j} = \sum_{\si \in \S_{r-1}}
\, \left[ g_{1 \si_1}^{(++)} +  g_{1 \si_1}^{(+-)} +
g_{1 \si_1}^{(-+)} + g_{1 \si_1}^{(--)}
 \right]
\, \otimes e_{j}^{(0)} \, e_{\si_j}^{(0)} \,  \om_{\si_2 \si_{j+1}}
\label{3-17}
\eeq
where $\si_{1}$ is fixed at $\si_{1} = 1$.
As we shall discuss later,
there are other exceptional indices that follow the expression (\ref{3-17});
a symmetry analysis of a Chan-Paton factor in a gravitational
holonomy operator would reveal that there are in fact two other such indices.
(In terms of the numbering indices $\si_{i}$ $(i = 1, 2, \cdots , n)$, the
exceptional ones can be given by $\si_{1} = 1$, $\si_{n-1} = n-1$ and $\si_{n} = n$.)

\noindent
\underline{Chan-Paton factors of gravitons}

From (\ref{3-10})-(\ref{3-17}), we can rewrite the comprehensive field $H$ as
\beqar
H &=& \sqrt{8 \pi G_{N}} \sum_{1 \le i < j \le r} \, \sum_{\si \in \S_{r-1}} \,
\bra E_{ij} \cdot E_{\si_{i} \si_{j}} \ket ~ \om_{ij} ~ \om_{\si_{i+1} \si_{j+1}}
\label{3-18} \\
\bra E_{ij} \cdot E_{\si_{i} \si_{j}} \ket &=&
\sum_{h_i} \, T^{\si_i} \, g_{i \si_i}^{(h_i)} \, \otimes \, e_{j}^{(0)} \, e_{\si_j}^{(0)}
~=~ \sum_{h_{i \si_i}} \, g_{i}^{(h_{i \si_i})} \, \otimes \, g_{j}^{(00)}
\label{3-19} \\
g_{i}^{(h_{i \si_i})} & \equiv & T^{\si_{i}} \, g_{i \si_i}^{(h_{i \si_i})}
\, = \, T^{\si_{i}} \, e_{i}^{(h_i) a} \, e_{\si_i}^{(h_{\si_i}) a}
\label{3-20}\\
g_{j}^{(00)} & \equiv & ( {\bf 1} )^{\si_j} \, e_{j}^{(0)} \, e_{\si_j}^{(0)}
\label{3-21}
\eeqar
where the sum of $h_{i \si_i}$ is taken over
$h_{i \si_i} \equiv h_{i} h_{\si_i} = (++, +-, -+, --)$.
The expression (\ref{3-19}) is analogous to that of (\ref{2-21}) with (\ref{2-26}).
Thus we can naturally interpret $T^{\si_i}$ as a Chan-Paton factor of
the graviton operators $g_{i}^{(++)}$ and $g_{i}^{(--)}$.
Notice that the color indices are now given by the numbering indices
to be permuted, which is natural since Chan-Paton factors of
gravitons should be encoded by braid generators.
We have not explicitly used the braid generators, however,
as discussed earlier, information of braid generators is
in one-to-one correspondence with permutations of the indices
as long as we use an irreducible representation.

In the above expressions,
the operators of $g_{i \si_i}^{(+-)}$ and $g_{i \si_i}^{(-+)}$
are naturally incorporated.
These represent massless spin-zero particles with no charges.
Life times of these should be the same as those of gravitons.
So they are stable.
We may therefore think of these spin-zero particles as a candidate for
the origin of dark matter or something that couples to dark energy.

We find that $H$ in the form of (\ref{3-18}) is
basically given by a sum over non-vanishing products of
the Chan-Paton factors of frame fields or
the translational operators in tangent spaces.
This is desirable in a view that
we need to integrate over all metrics for the construction
metric-free or topological theories.
Notice that we also need to have another sum over the
permutations so as to take care of diffeomorphism invariance.
In the construction of a gravitational theory, this is
realized by a braid trace in a holonomy operator.
This provides another reason for considering a holonomy formalism
in search of a theory of gravity.
As mentioned elsewhere, a full definition of $H$ is
then clarified after the construction and computation of
a gravitational holonomy operator. (See the expression (\ref{4-7})
for the full definition of $H$.)

\noindent
\underline{Summary}

In this section, we consider a comprehensive graviton field $H$ in a
framework of the holonomy formalism
which we have introduced in the previous section.
Since we have not yet constructed a holonomy operator of $H$,
a full-fledged definition of $H$ is to be obtained in the next section.
The results of the present section are, however,
quintessential for the construction of gravitational theories in twistor space.
These results can be summarized as follows.
\begin{enumerate}
  \item Diffeomorphism invariance in the spinor-momenta formalism is
  realized by a braid trace or a sum over permutations of the numbering indices.
  \item A Chan-Paton factor of a frame field is given by a translational operator or
  a four-momentum in a tangent space.
  \item A physical configuration space of gravitons is given by $\C \times \C$, where
  $\C = {\bf C}^n / \S_n$. Here ${\bf C}^n$ denotes complex number
  and $\S_n$ denotes the rank-$n$ symmetric group.
  \item Accordingly, a quantum Hilbert space of gravitons is given by
  $V^{\otimes n} \otimes V^{\otimes n}$,
  where $V^{\otimes n} = V_1 \otimes V_2 \otimes \cdots \otimes V_n$, with
  $V_i$ ($i= 1, 2, \cdots , n$) representing a Fock space that one set of the
  frame field operators $e_{i}^{(\pm)}$ act on. The other half of the Hilbert
  space consists of Fock spaces for another set of frame fields $e_{\si_i}^{(\pm)}$.
  \item Even though the Fock space of $e_{i}^{(\pm)}$ and that
  of $e_{\si_i}^{(\pm)}$ are different from each other,
  tangent spaces of the two operators are common since
  these are the constituents of a single graviton operator.
  A product of their Chan-Paton factors is defined on the
  common tangent space and is interpreted as a Chan-Paton factor of the graviton.
  \item  How to choose couplings of translational operators in tangent spaces
  is determined by braid generators.
  In other words, a Chan-Paton factor of a single graviton is encoded by
  a permutation of the numbering indices that label gravitons.
  An explicit form of this factor is essentially given by $T^{\si_i}$ in (\ref{3-12}).
  For a full definition, we also need $T^{\tau_i}$ in (\ref{4-3}) to be
  defined in the next section.
  \end{enumerate}

\section{Gravitational holonomy operators}

In analogy with the Yang-Mills case (\ref{2-25}), we
can construct a holonomy operator of the comprehensive
graviton field $H$ as
\beq
\Theta_{R, \ga}^{(H)} (u, \bu) = \Tr_{R, \ga} \, \Path \exp \left[
\sum_{m \ge 5} \oint_{\ga} \underbrace{H \wedge H \wedge \cdots \wedge H}_{m}
\right]
\label{4-1}
\eeq
where $\ga$ represents a closed path on $\C$ along which the integral is
evaluated and $R$ denotes representations of Poincar\'{e} algebra
and Iwahori-Hecke algebra.
Since $H$ is defined on $\C \times \C$, the integral should be interpreted
as a double integral.
The loop $\ga$ is commonly defined on each of $\C$'s.
This point will be clearer by the end of this section.
The condition $m \ge 5$ will also be clarified later.
As in the Yang-Mills case, the symbol $\Path$ denotes
the ordering of the numbering indices.
As discussed in the previous section,
Chan-Paton factors of $H$ depend on a permutation of the numbering indices.
In practical computations, we need to clarify this dependence.
Thus, in the following, we consider an exact meaning of
the Chan-Paton factor in (\ref{4-1}).

\noindent
\underline{Symmetries of Chan-Paton factors}

We first consider the significance of the fact that
the Chan-Paton factors of the comprehensive frame fields have vectorial properties.
In the Yang-Mills case, the Chan-Paton factor has a cyclic property
and we relate this to a $U(1)$ symmetry of the factor.
In a gravitational case, the relevant symmetry can
be given by a symmetry for a set of translational operators in tangent spaces.
A tangent space is a copy of a coordinate space
and the translational operators or the four-momenta can be encoded by the spinor momenta.
Thus the relevant symmetry is given by a symmetry for a set of the spinor momenta.
Since these spinor momenta are defined on $\cp^1$, we can map their
local coordinates $z_i$ ($i = 1, 2, \cdots , m$) on a complex $z$-plane
by stereographic projections.
It is well known that the conformal transformations of the complex $z$-plane
(including the point at infinity) is given by an $SL(2, {\bf C})$ group.
Thus, apart from gauge symmetries related to
Pincar\'e algebra and Iwahori-Hecke algebra,
we can identify the symmetry of the gravitational Chan-Paton factor
as the $SL(2, {\bf C})$ symmetry.
This symmetry is also related to the Lorentz invariance of
$u_i$'s as shown in (\ref{2-3}).

We now consider the effects of the $SL(2, {\bf C})$ symmetry on the braid trace.
First of all, we note that the number of elements in the
$SL(2, {\bf C})$ group is three.
This suggests that there exist three indices which characterize
the braid trace.
Remember that in the Yang-Mills case the Chan-Paton factor
has been characterized by one index due to the $U(1)$ symmetry
of the Chan-Paton factor.
The index $1$ has been chosen for a fixed numbering index for this reason.
This also corresponds to the fact that we have a single
closed loop along which a braid trace is defined.
Notice that mathematically it is known that a loop (or a link) forms
a braid group under {\it isotopy} of the loop.
In this sense, the loop can be denoted as $\ga_1$.
If we have an $SL(2, {\bf C})$ symmetry for the Chan-Paton factor
as in the present case, then the braid trace is characterized by
three distinct loops or three disconnected links.
Following a convention, we can choose the
three numbering indices as $(1, m-1, m)$ so that corresponding loops are
labeled by $\ga_1$, $\ga_{m-1}$, $\ga_m$.
This means that we have permutations of the numbering elements
$\{ 2,3,\cdots, m-2 \}$ in the definition of the holonomy operator (\ref{4-1}).

\begin{figure} [htbp]
\begin{center}
\includegraphics[width=140mm]{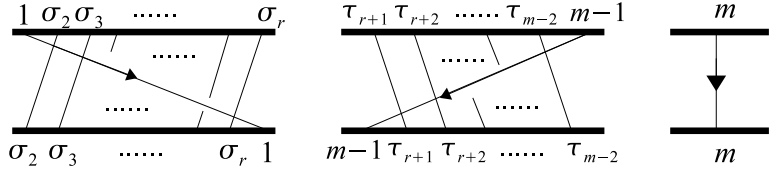}
\caption{Braid diagrams for the calculation of Chan-Paton factors ---
to make a loop out of each diagram, we need to connect an index on the top
with the one at the vertical bottom. The starting points
can be chosen as the top indices $(1, \, m-1, \, m)$, with orientations of
loops shown by arrows ($m \ge 5$).
When two lines are crossing each other, we consider that a line with
an arrow is closer to us, crossing over the other line without an arrow.}
\label{fig01}
\end{center}
\end{figure}

There must be correspondence between the
loops $( \ga_1 , \ga_{m-1} , \ga_m )$ and
the elements of $SL(2, {\bf C})$ algebra, say, a set of
generators $( t^{(+)} , t^{(-)} , t^{(0)} )$.
Since the $SL(2, {\bf C})$ symmetry is global or
comprehensive in the present context,
these generators are not labeled by a particular numbering index.
Instead, a set of numbering indices can be used to define
a ``state'' of a loop which is characterized by each of the generators.
Such a characterization can be carried out as follows.
We first regard the numbering index as something analogous to
a quantum number of $z$-direction in the conventional angular momentum algebra.
In terms of this number, the loop $\ga_m$ which corresponds to $t^{(0)}$ is trivial.
The loop $\ga_m$ is then expected to have only one element
for the numbering index, otherwise the $SL(2, {\bf C})$ symmetry
would be enhanced to include more $U(1)$ symmetries.
Thus, basically, distinct loops of the Chan-Paton factor is
characterized by the ladder generators $t^{(\pm)}$ of $SL(2, {\bf C})$.
One natural way of realizing this characterization
is to make the assigned elements of numbering indices
in a descending order for the loop $\ga_{1}$ and in an ascending order
for the loop $\ga_{m-1}$, along with certain orientations of the loops.
Let us denote the elements of $\ga_1$ by $\{\si_2 , \si_3 , \cdots ,
\si_{r} \}$ and those of $\ga_{m-1}$ by $\{ \tau_{r+1} , \tau_{r+2} ,
\cdots , \tau_{m-2} \}$ ($2 \le r \le m-3 $).
Then the three disconnected loops can
diagramatically be shown as Figure \ref{fig01}.
In the figure, the numbering elements are ordered by
$\si_2 < \si_3 < \cdots < \si_{r}$ and $\tau_{r+1} < \tau_{r+2}
< \cdots < \tau_{m-2}$, with the union of these elements
being $\{ 2, 3, \cdots , m-2 \}$.

Notice that an ordering of the numbering elements
naturally arises from the characterization of loops
by the $SL(2, {\bf C})$ algebra.
This is a supportive fact for the appearance of
the ordering symbol $\Path$ in the holonomy operator (\ref{4-1}).

\noindent
\underline{Explicit calculations}

By use of the above analysis, we now calculate a Chan-Paton factor
of the following quantity.
\beq
\Tr_{R, \ga} \, \Path \oint_{\ga}
\underbrace{H \wedge H \wedge \cdots \wedge H}_{m}
\label{4-2}
\eeq
This is essentially the exponent of the holonomy operator (\ref{4-1});
to obtain a full form, we simply take a sum over $m \ge 5$.
Probably, the simplest calculation is given by
making an assignment of the numbering elements
$\{2,3,\cdots, r\}$ to $\si$'s and $\{r+1, r+2 , \cdots , m - 2 \}$
to $\tau$'s, respectively.
So the numbering elements are split into two parts.
Under the ordering conditions,
$\si_2 < \si_3 < \cdots < \si_{r}$ and $\tau_{r+1} < \tau_{r+2}
< \cdots < \tau_{m-2}$, these elements are uniquely determined.
There is another way of calculating the Chan-Paton factor in (\ref{4-2}).
This can be carried out by assigning $\si$'s and $\tau$'s
to the overall elements $\{ 2, 3, \cdots , m-2 \}$ homogeneously.
Namely, the elements of both $\si$'s and $\tau$'s can take any values in
the overall elements, given that they satisfy the ordering conditions.
In the present paper, we shall leave this homogeneous case aside and
consider that split case only.\footnote{
Consideration of the homogeneous case will give an interpretation of
$\Theta_{R, \ga}^{(H)}$ as a square of $\Theta_{R, \ga}^{(E)}$ in
an intriguing way \cite{Abe:2005se}.
This point will be investigated in a separate paper.}

There are essentially two important ingredients in an explicit calculation
of the Chan-Paton factor, which can be stated as follows.
\begin{enumerate}
  \item A sum over all possible metrics: This is necessary for the
  construction of a gravitational theory which preserves general covariance.
  \item A braid trace or a sum over permutations of the numbering elements:
  This is necessary for diffeomorphism invariance.
\end{enumerate}
In the present case, we have two independent permutations, {\it i.e.},
$\si=\left(%
\begin{array}{c}
  2 \cdots r \\
  \si_2 \cdots \si_r \\
\end{array}%
\right)$
and
$\tau=\left(%
\begin{array}{c}
  r+1 \cdots m-2 \\
  \tau_{r+1} \cdots \tau_{m-2} \\
\end{array}%
\right)$.
A sum over all possible metrics
is then realized by a sum over these two permutations combined.
On the other hand, as in the Yang-Mills case, a braid trace can be realized by
a sum over a permutation of the overall elements $\{ 2, 3, \cdots , m-2 \}$.
This sum (or trace) should be taken on top of the sum over
the permutations of $\si$'s and $\tau$'s, which
suggests that the Chan-Paton factor in (\ref{4-2}) is
independent of the choice of $r$ ($2 \le r \le m-3$).

\begin{figure} [htbp]
\begin{center}
\includegraphics[width=140mm]{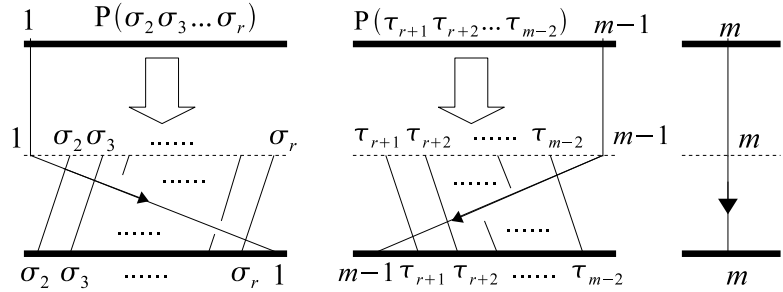}
\caption{Braid diagrams that include permutations of $\si$'s and $\tau$'s
--- the symbol $\Path$ denotes an ascending ordering of the arguments.
A sum over all permutations corresponds to a sum over all possible metrics
in a multigraviton system.}
\label{fig02}
\end{center}
\end{figure}

In Figure \ref{fig02}, we show braid diagrams that take account of
the permutations of $\si$'s and $\tau$'s.
In the figure, the elements of $\si$'s and $\tau$'s
are in a random order, while the symbol $\Path$ denotes an
ascending ordering of the elements.
A graviton labeled by a particular
numbering index corresponds to a particular strand in the braid diagrams.
A Chan-Paton factor of a graviton for a specific choice
of the permutations can then be encoded by corresponding braid diagrams.
Structures of the stands are schematically
shown by thick down-arrows in Figure \ref{fig02}.
As discussed in the previous section, explicit forms of
Chan-Paton factors for gravitons are determined by these structures.
A diagram on the left side in Figure \ref{fig02}
is the same as the one in Figure \ref{fig}.
Graviton Chan-Paton factors pertinent to this
diagram are therefore given by (\ref{3-12}).
Similarly, we can define graviton Chan-Paton factors
pertinent to a diagram in the center of Figure \ref{fig02} as
\beqar
\nonumber
T^{\tau_{i}} &=&
\left\bra p_{\tau_i}  \cdot (p_{m-1} + p_{\tau_{i < r+1}} + p_{\tau_{i < r+2}} +
\cdots + p_{\tau_{i < i -1 }} )  \right\ket
\\
&=& \left\bra
p_{\tau_i} \cdot \left( p_{m-1} + \sum_{k = r+1}^{i-1} p_{\tau_{i < k}} \right)
\right\ket ~~~~~~ \mbox{for $\, i=r+1, r+2, \cdots , m-2$}
\label{4-3}\\
T^{\tau_{m-1}} &=& T^{m-1} ~ = ~ 1
\label{4-4}\\
p_{\tau_{i<j}} & \equiv &  \left\{
    \begin{array}{ll}
      p_{\tau_i} & \mbox{for $\tau_i < \tau_j$}\\
      0  & \mbox{otherwise}
    \end{array} \right.
\label{4-5}
\eeqar
where the definition of $p_{\tau_{i < j}}$ is the same as (\ref{3-13})
except that we replace $\si_i$ ($i = 2, 3, \cdots, r$) by
$\tau_{i}$ ($i= r+1 , r+2 , \cdots , m-2$).
A graviton Chan-Paton factor corresponding to
a diagram on the right in Figure \ref{fig02} is simple.
Since there are no permutations involved, as in
the cases of $i = 1$ and $i= m-1$, the Chan-Paton factor of the
$m$-th graviton is defined as
\beq
T^{m} ~ = ~ 1
\label{4-6}
\eeq

Using these expressions, we now obtain a full form of
the comprehensive graviton field $H$ as follows.
\beqar
H &=& \sqrt{8 \pi G_N}
\sum_{1 \le i < j \le m} \,
\sum_{\si \in \S_{r-1}}
\sum_{\tau \in \S_{m-r-2}}
\left(
\sum_{h_i} g_{i}^{(h_{i \mu_i})} \otimes g_{j}^{(00)}
\right)
\, \om_{ij} \, \om_{\la_{i} \la_{j}}
\label{4-7} \\
\mu_i & = &  \left\{
    \begin{array}{ll}
      \si_{i} & \mbox{for $i = 1, 2, \cdots , r$}\\
      \tau_{i}  & \mbox{for $i = r+1, r+2, \cdots, m-1$} \\
      m  & \mbox{for $i = m$} \\
    \end{array} \right.
\label{4-8}\\
\la_i & = &  \left\{
    \begin{array}{ll}
      \si_{i+1} & \mbox{for $i = 1, 2, \cdots , r$}\\
      \tau_{i-1}  & \mbox{for $i = r+1, r+2, \cdots, m-1$} \\
      m  & \mbox{for $i = m$} \\
    \end{array} \right.
\label{4-9} \\
g_{i}^{(h_{i \mu_i})} & = &   T^{\mu_i} \, g_{i \mu_i}^{(h_{i \mu_i})}
\, = \, T^{\mu_i} \, e_{i}^{(h_{i}) a} \, e_{\mu_i}^{(h_{\mu_i}) a}
\label{4-10} \\
g_{j}^{(00)} & = &  ( {\bf 1} )^{\mu_j} \, e_{j}^{(0)} \, e_{\mu_j}^{(0)}
\label{4-11}
\eeqar
where the sum of $h_{i \si_i}$ is taken over
$h_{i \mu_i} \equiv h_{i} h_{\mu_i} = (++, +-, -+, --)$ as in (\ref{3-19}).
Notice that the index $\mu_i$ is a composite numbering index and that it should
not be confused with a Minkowski index.
From an $SL(2, {\bf C})$ symmetry of the
comprehensive graviton field, we can fix the following indices.
\beqar
\nonumber
&& \si_1 = 1 \, , ~ \tau_{m-1} = m-1 \, , ~ \si_{m} = m \, ,
\\
&& \la_{r} = 1 \, , ~ \la_{r+1} = m-1 \, , ~ \la_{m+1} = \si_2
\label{4-12}
\eeqar
These choices are in accord with the braid diagrams in Figure \ref{fig02}.
Information of $\la_{m+1}$ is necessary in defining the gravitational
holonomy operator (\ref{4-1}).
Rigorously speaking, the number of gravitons should be
represented by $n$ rather than $m$ in (\ref{4-7}).
If we substitute $H$ into the holonomy operator
$\Theta_{R, \ga}^{(H)} (u, \bu)$, the number $n$ effectively
becomes $m$ in the computations of the quantity (\ref{4-2}).
Thus, the expression (\ref{4-7}), along with
(\ref{3-12})-(\ref{3-14}), (\ref{4-3})-(\ref{4-6}) and (\ref{4-8})-(\ref{4-12}),
provides a full definition of the comprehensive fields for gravitons,
complementing the arguments in the previous section.

Our particular choice of the graviton Chan-Paton factors,
{\it i.e.}, $T^{\mu_i}$'s in (\ref{4-10}),
are determined by the braid diagrams in Figure \ref{fig02}.
These diagrams correspond an arbitrary
permutation of the numbering indices which respects
the $SL(2, {\bf C})$ symmetry of the Chan-Paton factor in the quantity (\ref{4-2}).
These interrelations arise from the fact that the Chan-Paton
factors of gravitons are made of the Poincar\'{e} algebra and the Iwahori-Hecke algebra.
Since Lorentz invariance is manifest in the holonomy formalism,
the Poincar\'{e} symmetry reduces to a symmetry of spacetime translations.
An irreducible representation of this symmetry
is given by translational operators or four-momenta, which we
have identified with the Chan-Paton factors of frame fields.
Needless to say, a quantum field theory is defined by
a unitary irreducible representation (UIR) of physical observables.
Irreducibility is crucial here to extract the pure four-momenta
as basic ingredients of the Chan-Paton factors.
The same argument applies to the Iwahori-Hecke algebra as well,
that is, the Chan-Paton factors of gravitons
should also be described by an irreducible representation of the braid generators.
In terms of the braid diagrams, irreducibility means
that the pattern of each diagram is uniquely determined up to isotopy or
the so-called Reidemeister moves.
There are in fact many irreducible representations in this regard.
Our choice of $T^{\mu_i}$'s is one of them.
In what follows, we shall see this point in a step-by-step manner,
starting from the case of $m = 5$ to more general cases.

\noindent
\underline{For $m = 5$}

As discussed before, the graviton Chan-Paton factors are characterized
by three distinct loops, due to an $SL (2, {\bf C})$ symmetry
of the Chan-Paton factors.
The condition $m \ge 5$ on the number of gravitons
is imposed by the very distinctiveness of the three loops.

\begin{figure} [htbp]
\begin{center}
\includegraphics[width=70mm]{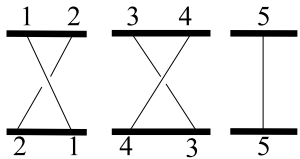}
\caption{Braid diagrams for $m=5$}
\label{fig03}
\end{center}
\end{figure}

For $m = 5$, there is only one element for either $\si$
or $\tau$, {\it i.e.}, $\si_{2} = 2$ or $\tau_{3} = 3$,
so that there are no permutations involved in braid diagrams.
This corresponds to the fact that there is only one choice
of $r$ ($2 \le r \le m-3$) for $m = 5$.
A structure of the diagrams is therefore uniquely
determined as in Figure \ref{fig03}.

In the case of $m = 5$,
the three indices to be fixed are given by $\{ 1, 4 , 5 \}$.
The graviton Chan-Paton factors labeled by these indices
are trivial, {\it i.e.}, $T^{1} = T^{4} = T^{5} = 1$.
Thus nontrivial factors arise from
the gravitons labeled by $\si_2 = 2$ and $\tau_{3} = 3$.
From (\ref{3-12}) and (\ref{4-3}), we find that these are given by
\beqar
\nonumber
f ( \si_2 ) & \equiv & T^{\si_2}  ~=~ \bra p_1 \cdot p_{\si_2} \ket  \\
\tilde{f} ( \tau_3 ) & \equiv & T^{\tau_3} ~=~ \bra p_{\tau_3} \cdot p_4 \ket
\label{4-13}
\eeqar
Notice that these can easily be read off from Figure \ref{fig03}.
We first look at the strand of $\si_2$ and then interpret the
crossing with the strand of $1$ as a coupling
between the Chan-Paton factors of the frame fields labeled by $\si_2$ and $1$.
A Chan-Paton factor relevant to the strand of $\tau_3$
can similarly be determined by the middle diagram in Figure \ref{fig03}.
Since there is no permutation, the diagrams are automatically irreducible
at this level.
It is, however, illustrative to express
the quantity (\ref{4-2}) for $m=5$ in terms of (\ref{4-13}).
We can write down an explicit expansion as
\beqar
\nonumber
&& \!\!\!\! \Tr_{R, \ga} \, \Path \oint_{\ga}
H \wedge H \wedge H \wedge H \wedge H
\\
\nonumber
&=& \left( 8 \pi G_N \right)^{\frac{5}{2}}
\Tr_{R , \ga} \oint_{\ga} H_{12} H_{23} \cdots H_{51}
\, \om_{12} \wedge \om_{23} \wedge \cdots \wedge \om_{51}
\\
\nonumber
&=&
\left( 8 \pi G_N \right)^{\frac{5}{2}}
\left( \frac{1}{2^{6}} \right)^2
f( \si_2 ) \, \tilde{f} ( \tau_3 )
\!\! \sum_{(h_{1 1} , h_{2 \si_2} , \cdots , h_{55})} \!\!
g_{1 1}^{(h_{1 1})} \otimes g_{2 \si_2}^{(h_{2 \si_2})} \otimes
g_{3 \tau_3}^{(h_{3 \tau_3})} \otimes g_{4 4}^{(h_{4 4})} \otimes
g_{5 5}^{(h_{5 5})}
\\
\nonumber
&&
\left. ~ \times
\oint_{\ga} \om_{ 1 2 } \wedge \om_{ 2 3 } \wedge
\om_{3 4} \wedge \om_{4 5} \wedge \om_{5 1}
\oint_{\ga} \om_{ \si_{2} 1} \wedge \om_{1 4} \wedge
\om_{4 \tau_3 } \wedge \om_{\tau_{3} 5} \wedge \om_{5 \si_{2} }
\right|_{\si_{2} = 2 , \, \tau_3 = 3}
\\
&&
~ + \, \P (2 3)
\label{4-14}
\eeqar
where the sum of $(h_{1 1} , h_{2 \si_2}, \cdots h_{55})$ is taken
over any combinations of $h_{i \mu_i} \equiv h_i h_{\mu_i} = (++, +- , -+ , --)$
for $i = 1,2, \cdots , 5$, with $\mu_i$ being defined as (\ref{4-8}).
$\P (2 3)$ in the last line indicates terms obtained by the permutation of
the numbering indices $\{ 2 , 3 \}$ or $\{ \si_2 , \tau_3 \}$.
This permutation arises from the braid trace $\Tr_{\ga}$.
As indicated in the second last line, $\si_2$ and $\tau_3$ are fixed.
This can be interpreted as non-existence of
a sum over possible metrics in the present case.
Such a sum appears for $m \ge 6$ as we shall see in the following.

\noindent
\underline{For $m = 6$}

\begin{figure} [htbp]
\begin{center}
\includegraphics[width=110mm]{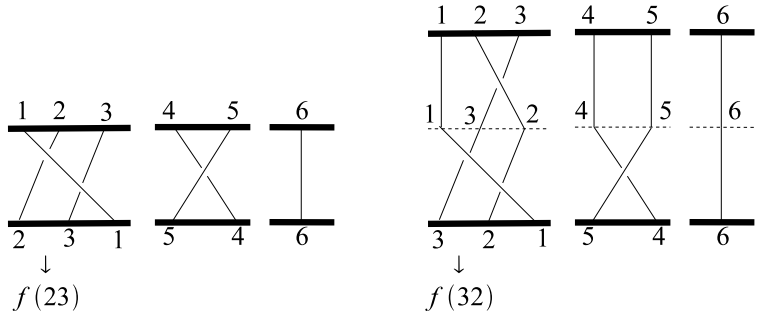}
\caption{Braid diagrams for $m=6$}
\label{fig04}
\end{center}
\end{figure}

In this case, the indices to be fixed are given by $\{ 1, 5, 6 \}$.
Those that are relevant to a braid trace are given by
$\{ \si_2 , \si_3 \} = \{ 2, 3 \}$ and $\tau_{4} = 4$.
The braid diagrams for the calculation of the Chan-Paton factor
in (\ref{4-2}) are then shown as Figure \ref{fig04}.

\begin{figure} [htbp]
\begin{center}
\includegraphics[height=60mm]{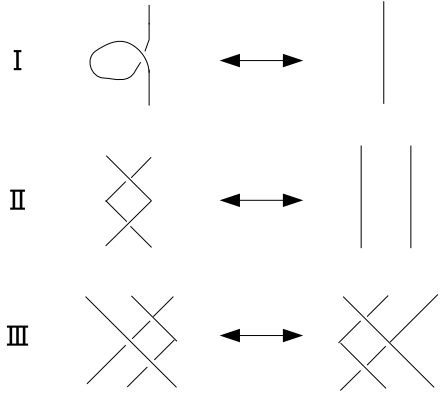}
\caption{Reidemeister moves corresponding to
the raising operator $t^{(+)}$ of the $SL(2, {\bf C})$ algebra}
\label{fig05}
\end{center}
\end{figure}

In analogy with (\ref{4-12}), we can define nontrivial factors
as $f(\si_{2} \si_{3})= T^{\si_{2}} T^{\si_{3}}$
and $\tilde{f}( \tau_4 ) = T^{\tau_{4}}$.
Explicit forms of these can be written as
\beqar
\nonumber
f(23) &=& \bra p_1 \cdot p_2 \ket \bra p_1 \cdot p_3 \ket
\\
f(32) &=& \bra p_1 \cdot p_2 \ket \bra ( p_1 + p_2 ) \cdot p_3 \ket
\label{4-15}
\\
\nonumber
\tilde{f}(4) &=& \bra p_4 \cdot p_5 \ket
\eeqar
By use of the previous rules, it is obvious that we can
reproduce these factors directly from Figure \ref{fig04}.
It is also easy to see irreducibility of the diagrams,
which can be understood as follows.
We first notice that the characterization of braid diagrams
in terms of an $SL (2, {\bf C})$ symmetry leads to
preservation of a basic pattern shown on the left side in Figure \ref{fig04}.
The right-hand side diagrams are build
on the basic pattern with a subdiagram that represents
a transposition of the indices $2$ and $3$.
These diagrams, which is essentially given by the one labeled by
$\{ 1,2,3 \}$, are uniquely determined up to isotopy.
In other words, the diagram of $\{ 1,2,3 \}$ in the right side of
Figure \ref{fig04} is irreducible up to the Reidemeister moves
shown in Figure \ref{fig05}.

\begin{figure} [htbp]
\begin{center}
\includegraphics[height=60mm]{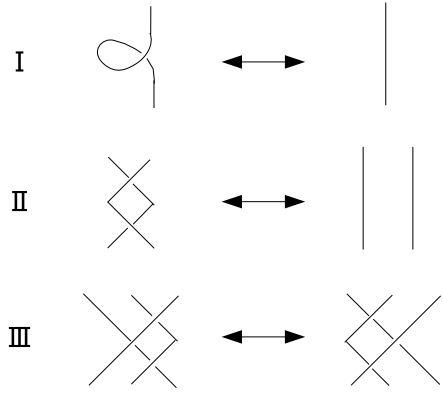}
\caption{Reidemeister moves corresponding to
the lowering operator $t^{(-)}$ of the $SL(2, {\bf C})$ algebra.}
\label{fig06}
\end{center}
\end{figure}

Notice that the diagrams labeled by the indices $\{ 1,2,3 \}$ correspond to
the raising operator $t^{(+)}$ of the $SL(2, {\bf C})$ algebra
and that they have a crossing rule, {\it i.e.},
if two strands are crossing each other, then
a strand with the smaller index crosses {\it over} the other strand.
We have actually drawn Figure \ref{fig}, following this rule.
On the other hand,
the diagrams labeled by the indices $\{ 4, 5 \}$ correspond
to the lowering operator $t^{(-)}$ of the $SL(2, {\bf C})$ algebra
and, in this case, a crossing rule can be stated as,
``If two strands are crossing each other, then
a strand with the larger index crosses {\it over} the other strand''.
The Reidemeister moves for this type of braid diagrams can be
given by Figure \ref{fig06}.

In both cases, the type-I moves are irrelevant in the present context.
Only the type-II and type-III moves will be used for the
irreducibility of the braid diagrams in general.
In the case of $m = 6$, there are only two numbering elements to be permuted.
Thus only the type-II move in Figure \ref{fig05} is used to obtain irreducible diagrams.
The type-III moves will be utilized for $m \ge 7$.

Without a notion of irreducibility, we can in principle put
any subdiagrams that are made of the Reidemeister moves
on top of the basic diagrams indicated in the left side
of Figure \ref{fig04}. Such procedures produce reducible diagrams.
In reducible diagrams, there are crossings between the same strands
more than once.
Thus irreducibility in this context means a fact that a particular
strand crosses with a specific strand once or none.
This condition is in fact satisfied for any $m$, and is reflected in
the definitions of $T^{\si_i}$ and $T^{\tau_i}$.

Using the expressions in (\ref{4-15}), we can explicitly calculate
the quantity (\ref{4-2}) as
\beqar
\nonumber
&& \!\!\!\! \Tr_{R, \ga} \, \Path \oint_{\ga}
H \wedge H \wedge H \wedge H \wedge H \wedge H
\\
\nonumber
&=& \!\! \left( 8 \pi G_N \right)^{3}
\Tr_{R , \ga} \oint_{\ga} H_{12} H_{23} \cdots H_{61}
~ \om_{12} \wedge \om_{23} \wedge \cdots \wedge \om_{61}
\\
\nonumber
&=& \!\!
\left( 8 \pi G_N \right)^{3}
\left( \frac{1}{2^{7}} \right)^{2} \sum_{\si \in \S_2 }
f( \si_2 \si_3 ) \, \tilde{f} ( \tau_4 )
\\
\nonumber
&&
~ \times
\! \sum_{(h_{1 1} , h_{2 \si_2}, \cdots h_{66})} \!
g_{1 1}^{(h_{1 1})} \otimes g_{2 \si_2}^{(h_{2 \si_2})} \otimes
g_{3 \si_3}^{(h_{3 \si_3})} \otimes g_{4 \tau_4}^{(h_{4 \tau_4})} \otimes
g_{5 5}^{(h_{5 5})} \otimes g_{6 6}^{(h_{6 6})}
\\
\nonumber
&& \left.
~ \times
\oint_{\ga} \om_{1 2} \wedge \om_{2 3} \wedge
\om_{3 4} \wedge \om_{4 5} \wedge \om_{5 6} \wedge \om_{6 1}
\oint_{\ga} \om_{ \si_{2} \si_{3}} \wedge \om_{\si_{3} 1} \wedge
\om_{1 5} \wedge \om_{5 \tau_{4}} \wedge \om_{\tau_{4} 6 } \wedge \om_{6 \si_{2}}
\right|_{\tau_{4} = 4 }
\\
&&
~ + \, \P (2 3 4)
\label{4-16}
\eeqar
Notice that we now have a sum over permutations of
$\si = \left(
    \begin{array}{c}
    2 ~ \, 3 \\
    \si_2 \, \si_3 \\
    \end{array}
\right)$.
This sum corresponds to a sum over possible metrics in a six-graviton system.

\noindent
\underline{For $m = 7$}

\begin{figure} [htbp]
\begin{center}
\includegraphics[width=120mm, height=90mm]{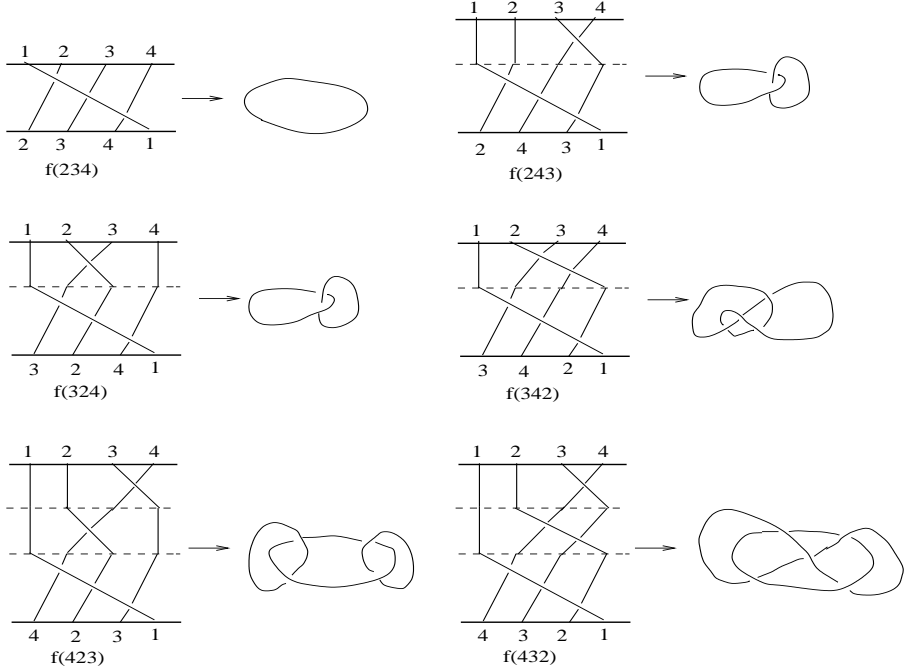}
\caption{Braid diagrams for $m=7$}
\label{fig07}
\end{center}
\end{figure}

At this stage, it is straightforward to extend our
formalism to the case of $m=7$.
We choose $r$ to be $r= 4$ so that the $\tau$-part of permutation
is trivially fixed at $\tau_5 = 5$.
The $\si$-part of the braid diagrams for $m=7$ are then given by Figure \ref{fig07}.

In Figure \ref{fig07}, a link which is associated with each of the
braid diagrams is also shown.
Notice that the last diagrams with the factor of $f(432)$
(at the right-bottom corner) contains a subdiagram that
is equivalent to the left-hand-side pattern of the type-III
Reidemeister moves in Figure \ref{fig05}.
Thus this subdiagram can be replaced by the
other pattern of the same type-III moves.
As is expected, such a replacement does not change the factor
of $f(432)$ or $T^{\si_i}$'s, which
shows another confirmation that $T^{\si_i}$'s correspond to
an irreducible representation of the braid generators.

So far, we have not made direct use of the braid generators.
This is because information of the braid generators is,
at the level of trace calculations, encoded by a permutation of
the numbering indices. An extraction of a specific braid generator
labeled by a single numbering element therefore does not lead
to physical quantities; we rather need information of
full or comprehensive permutations of the indices in order to obtain
physical quantities.
In the present case, such information is given by
$f (\si_2 \si_3 \si_4 )$, which can be expressed as
\beq
f (\si_2 \si_3 \si_4 )
\, = \, T^{\si_2} T^{\si_3} T^{\si_4}
\, \equiv \, \Tr_{R} \oint_{\ga_1} B_{\si_2 \si_3 \si_4}
\label{4-17}
\eeq
where we introduce notation $B_{\si_2 \si_3 \si_4}$
to indicate dependence on braid generators.
An irreducible representation of the braid generators
are given by the elements of Iwahori-Hecke algebra $\tilde{b}_i$ in (\ref{2-19}).
These elements depend on the numeric factor $\eta = \exp ( i \pi / 2 \kappa)$.
Thus we expect some contributions of this factor $\eta$ in $B_{\si_2 \si_3 \si_4}$;
clarification of this point is currently under investigation.

In (\ref{4-17}), $\Tr_{R}$ denotes a trace over Poincar\'{e} algebra.
This trace also implies a fact that we can actually use any index
among $\{ 1, \si_2 , \si_3 , \si_4 \}$ as a representing index
of a loop corresponding to the raising operator of the $SL (2 ,{\bf C})$
algebra. Of course, this trace does not mean cyclicity of
the indices at all.
In fact, as in the previous cases, explicit forms of (\ref{4-17})
can easily be obtained from the definition of $T^{\si_i}$'s as
\beqar
\nonumber
f(234) ~=~ \Tr_{R} \oint_{\ga_1} B_{234} &=& \bra p_1 \cdot p_2 \ket
\bra p_1  \cdot p_3 \ket \bra p_1 \cdot p_4 \ket
\\ \nonumber
f(243) ~=~ \Tr_{R} \oint_{\ga_1} B_{243} &=& \bra p_1 \cdot p_2 \ket
\left\bra (p_1 + p_3 ) \cdot p_4 \right\ket \bra p_1 \cdot \p_3 \ket
\\ \nonumber
f(324) ~=~ \Tr_{R} \oint_{\ga_1} B_{324} &=& \left\bra ( p_1 + p_2 ) \cdot p_3 \right\ket
\bra p_1 \cdot p_2 \ket \bra p_1 \cdot p_3 \ket
\\
f(342) ~=~ \Tr_{R} \oint_{\ga_1} B_{342} &=& \left\bra ( p_1 + p_2 ) \cdot p_3 \right\ket
 \left\bra ( p_1 + p_2 ) \cdot p_4 \right\ket  \bra p_1  \cdot p_2 \ket
\label{4-18}
\\ \nonumber
f(432) ~=~  \Tr_{R} \oint_{\ga_1} B_{432} &=& \left\bra ( p_1 + p_2 + p_3 ) \cdot p_4 \right\ket
\left\bra ( p_1 + p_2 )\cdot p_3 \right\ket \bra p_1 \cdot p_2 \ket
\\ \nonumber
f(423) ~=~  \Tr_{R} \oint_{\ga_1} B_{423} &=& \left\bra ( p_1 + p_2 + p_3 ) \cdot p_4 \right\ket
\left\bra p_1 \cdot p_2 \right\ket \bra p_1 \cdot p_3 \ket
\eeqar
In terms of these factors, the quantity (\ref{4-2}) can be calculated as
\beqar
\nonumber
&& \!\!\!\! \Tr_{R, \ga} \, \Path \oint_{\ga}
H \wedge H \wedge H \wedge H \wedge H \wedge H \wedge H
\\
\nonumber
&=& \!\! \left( 8 \pi G_N \right)^{\frac{7}{2}}
\Tr_{R , \ga} \oint_{\ga} H_{12} H_{23} \cdots H_{71}
~ \om_{12} \wedge \om_{23} \wedge \cdots \wedge \om_{71}
\\
\nonumber
&=& \!\!
\left( 8 \pi G_N \right)^{\frac{7}{2}}
\left( \frac{1}{2^{8}} \right)^{2} \sum_{\si \in \S_3 }
f( \si_2 \si_3 \si_4 ) \, \tilde{f} ( \tau_4 )
\\
\nonumber
&& ~ \times \! \sum_{(h_{11} , h_{2 \si_2} , \cdots , h_{77} )} \!
g_{1 1}^{(h_{11})} \otimes g_{2 \si_2}^{(h_{2 \si_2})} \otimes
g_{3 \si_3}^{(h_{3 \si_3})} \otimes g_{4 \si_4}^{(h_{4 \si_4})} \otimes
g_{5 \tau_5 }^{(h_{5 \tau_5})} \otimes g_{6 6}^{(h_{66})} \otimes g_{7 7}^{(h_{77})}
\\
\nonumber
&&
~ \times
\oint_{\ga} \om_{ 1 2 } \wedge \om_{ 2 3 } \wedge
\om_{3 4} \wedge \om_{4 5} \wedge \om_{5 6} \wedge \om_{6 7} \wedge \om_{7 1}
\\
\nonumber
&& ~ \times
\left.
\oint_{\ga} \om_{ \si_{2} \si_{3}} \wedge \om_{\si_{3} \si_{4}} \wedge \om_{\si_{4} 1}
\wedge \om_{1 6} \wedge \om_{6 \tau_{5}} \wedge \om_{\tau_{5} 7 } \wedge \om_{7 \si_{2}}
\right|_{\tau_{5} = 5 }
\\
&&
~ + \, \P (2 3 4 5)
\label{4-19}
\eeqar

\noindent
\underline{General cases}

For completion of the discussion, in the following
we present an explicit calculation of the quantity (\ref{4-2}) for arbitrary $m$.
\beqar
\nonumber
&& \!\!\!\! \Tr_{R, \ga} \, \Path \oint_{\ga}
\underbrace{H \wedge H \wedge \cdots \wedge H}_{m}
\\
\nonumber
&=& \!\! \left( 8 \pi G_N \right)^{\frac{m}{2}}
\Tr_{R , \ga} \oint_{\ga} H_{12} H_{23} \cdots H_{m1}
~ \om_{12} \wedge \om_{23} \wedge \cdots \wedge \om_{m1}
\\
\nonumber
&=& \!\!
\left( 8 \pi G_N \right)^{\frac{m}{2}}
\left( \frac{1}{2^{m+1}} \right)^{2} \sum_{\si \in \S_{r-1} }
\sum_{\tau \in \S_{m-r-2} }
f( \si ) \, \tilde{f} ( \tau )
\\
\nonumber
&& ~ \times \! \sum_{(h_{11} , h_{2 \si_2} , \cdots , h_{m m} )} \!
g_{1 1}^{(h_{11})} \otimes g_{2 \si_2}^{(h_{2 \si_2})} \otimes
g_{3 \si_3}^{(h_{3 \si_3})} \otimes \cdots \otimes g_{r \si_r}^{(h_{r \si_r})}
\\
\nonumber
&& \hskip 2.7cm
\otimes \, g_{r+1 \, \tau_{r+1} }^{(h_{r+1 \, \tau_{r+1}})}
\otimes g_{r+2 \, \tau_{r+2} }^{(h_{r+2 \, \tau_{r+2}})}
\otimes \cdots \otimes g_{m-2  \, \tau_{m-2} }^{(h_{m-2 \, \tau_{m-2}})}
\otimes g_{m-1 \, m-1}^{(h_{m-1 \, m-1})} \otimes g_{m m}^{(h_{m m})}
\\
\nonumber
&& ~ \times
\oint_{\ga} \om_{ 1 2 } \wedge \om_{ 2 3 } \wedge \cdots \wedge \om_{m-1 \, m}
\wedge \om_{m 1}
\\
\nonumber
&& ~ \times
\oint_{\ga} \om_{ \si_{2} \si_{3}} \wedge \om_{\si_{3} \si_{4}} \wedge \cdots \wedge
\om_{\si_{r-1} \si_{r}} \wedge \om_{\si_{r} 1}
\\
\nonumber
&& \hskip 3.8cm
\wedge \, \om_{1 \, m-1}  \wedge \om_{m-1 \, \tau_{r+1}} \wedge \om_{\tau_{r+1} \tau_{r+2}}
\wedge \cdots \wedge \om_{\tau_{m-2} \, m } \wedge \om_{m \si_{2}}
\\
&&
~ + \, \P (2 3 \cdots m-2)
\label{4-20}
\eeqar
where $f(\si)$ and $\tilde{f} (\tau)$ are defined as
\beq
f (\si) = \prod_{i=2}^{r} T^{\si_i} \, , ~~~
\tilde{f} (\tau) = \prod_{i=r+1}^{m-2} T^{\tau_i}
\label{4-21}
\eeq
Explicit forms of $T^{\si_i}$'s and $T^{\tau_i}$'s are defined in
(\ref{3-12}) and (\ref{4-3}), respectively.

In (\ref{4-20}), a sum over possible metrics is realized by
the double sum over the permutations of
$\si=\left(%
\begin{array}{c}
  2 \cdots r \\
  \si_2 \cdots \si_r \\
\end{array}%
\right)$
and
$\tau=\left(%
\begin{array}{c}
  r+1 \cdots m-2 \\
  \tau_{r+1} \cdots \tau_{m-2} \\
\end{array}%
\right)$.
On the other hand, a braid trace is realized by
$\P(23 \cdots m-2)$, which indicates
the terms obtained by permutations of the overall elements
$\{2,3, \cdots, m-2 \}$.


An explicit description of the gravitational holonomy operator (\ref{4-1})
in terms of the graviton operator $g_{i \mu_i}^{(h_i)}$,
where $\mu_i = ( \si_i , \tau_i)$ denotes a composite index,
can then be given by the expression (\ref{4-20}).
We consider that the gravitational holonomy operator defines
a theory of gravity in twistor space and that any physical
quantities, such as graviton amplitudes, are generated from  this holonomy operator.
Indeed, the structure of the Chan-Paton factor in (\ref{4-20}) is the same as that of
graviton amplitudes which has been obtained by Bern et al. in \cite{Bern:1998sv}.
In fact, the definition of $T^{\si_i}$ in (\ref{3-12}) or $T^{\tau_i}$
in (\ref{4-3}) is inspired by the results of \cite{Berends:1988zp, Kawai:1985xq}
and \cite{Bern:1998sv}.
In the next section, we shall use these relations to obtain an S-matrix functional of
graviton amplitudes in terms of a supersymmetric version of
the gravitational holonomy operator.

\section{An S-matrix functional for graviton amplitudes}

In this section, we obtain an S-matrix functional for graviton amplitudes,
following a case of gluon amplitudes discussed
in the accompanying paper \cite{Abe:hol01}.
For this purpose, we first review an $\N = 4$ supersymmetric extension
of the Yang-Mills holonomy operator (\ref{2-25}) and how it can be used
to describe an S-matrix functional for gluon amplitudes.
(This review part can be omitted
if the reader is already familiar with the material in \cite{Abe:hol01}.)
We then apply these results to a gravitational theory which we have
formulated in the previous two sections.

\noindent
\underline{Supersymmetrization of $\Theta_{R, \ga}^{(A)} (u)$}

In the following, we simply present some key results in \cite{Abe:hol01}.
A supersymmetric extension of (\ref{2-25}) can be expressed as
\beq
\Theta_{R, \ga}^{(A)} (u; x, \th) = \Tr_{R, \ga} \, \Path \exp \left[
\sum_{m \ge 2}^{\infty} \oint_{\ga} \underbrace{A \wedge A \wedge \cdots \wedge A}_{m}
\right]
\label{5-1}
\eeq
where, as in (\ref{2-20})-(\ref{2-21}), $A$ is defined by
\beqar
A &=&  g \sum_{1 \le i < j \le n} A_{ij} \, \om_{ij}
\label{5-2}\\
A_{ij} &=& \sum_{\hat{h}_i} a_{i}^{(\hat{h}_{i})} (x, \th) \otimes a_{j}^{(0)}
\label{5-3} \\
\om_{ij} & = & d \log(u_i u_j) = \frac{d(u_i u_j)}{(u_i u_j)}
\label{5-4}
\eeqar
These expressions are the same as the previous ones except that
physical operators $a_{i}^{(\hat{h}_{i})}$ are now dependent on
the four-dimensional chiral supercoordinate $(x, \th)$.
Accordingly, the physical operators include the states of
gluonic superpartners, so that the helicity index is
extended from $h_i$ to $\hat{h}_{i}$ which we shall specify in a moment.
In the Yang-Mills case, we consider $\N = 4$ supersymmetry.
So $\th$ is written as $\th_{A}^{\al}$ $(A = 1,2; \al = 1,2,3,4)$.
Projection of these Grassmann variables onto a $\cp^1$ fiber of supertwistor space
is realized by
\beq
\xi^\al = \th_{A}^{\al} u^A
\label{5-5}
\eeq
In terms of these, an explicit form of $a_{i}^{(\hat{h}_{i})} (x, \th)$ is given by
\beq
a_{i}^{(\hat{h}_{i})} (x, \th)
\, = \,
\left. \int d\mu (p_i) ~ a_{i}^{(\hat{h}_{i})} (\xi_i) ~  e^{ i x_\mu p_{i}^{\mu} }
\right|_{\xi_{i}^{\al} = \th_{A}^{\al} u_{i}^{A} }
\label{5-6}
\eeq
where $a_{i}^{(\hat{h}_{i})} (\xi_i)$'s are defined as
\beqar
\nonumber
a_{i}^{(+)} (\xi_i) &=& a_{i}^{(+)} \\ \nonumber
a_{i}^{\left( + \hf \right)} (\xi_i) &=& \xi_{i}^{\al}
\, a_{i \al}^{ \left( + \hf \right)} \\
a_{i}^{(0)} (\xi_i) &=& \hf \xi_{i}^{\al} \xi_{i}^{\bt} \, a_{i \al \bt}^{(0)} \label{5-7}
\\ \nonumber
a_{i}^{\left(- \hf \right)} (\xi_i) &=& \frac{1}{3!} \xi_{i}^{\al}\xi_{i}^{\bt}\xi_{i}^{\ga}
\ep_{\al \bt \ga \del} \, {a_{i}^{ \del}}^{ \left( - \hf \right)}
\\ \nonumber
a_{i}^{(-)} (\xi_i) &=& \xi_{i}^{1} \xi_{i}^{2} \xi_{i}^{3} \xi_{i}^{4} \, a_{i}^{(-)}
\eeqar
Notice that the helicity components are in accordance with the relation in (\ref{2-9}).
The measure $d \mu (p)$ in (\ref{5-6}) denotes the following Lorentz invariant measure.
\beqar
\nonumber
d \mu (p) \equiv
\frac{d^3 p}{(2 \pi)^3} \frac{1}{2 p_0}
&=&
\frac{1}{(2 \pi)^3} \frac{({\bar \al} \al) d ({\bar \al} \al)}{2} \frac{ dz d \bz}{(-2i)}
\\
&=&
\frac{1}{4} \left[
\frac{u \cdot du}{2 \pi i} \frac{d^2 \bu}{(2 \pi)^2} -
\frac{\bu \cdot d \bu}{2 \pi i} \frac{d^2 u}{(2 \pi)^2}
\right]
\label{5-8}
\eeqar
This is called the Nair measure.

\noindent
\underline{An S-matrix functional for gluon amplitudes}

In the spinor-momenta formalism, the simplest way of
describing gluon amplitudes is to factorize the amplitudes
in terms of the maximally helicity violating (MHV) amplitudes.
The MHV amplitudes are the scattering amplitudes of $(n-2)$
positive-helicity gluons and $2$ negative-helicity gluons.
In a momentum-space representation, the MHV tree amplitudes
are expressed as
\beqar
\nonumber
\A_{MHV}^{(1_+ 2_+ \cdots r_{-} \cdots s_{-} \cdots n_+ )} (u, \bu)
& \equiv &
\A_{MHV}^{(r_{-} s_{-})} (u, \bu) \\
& = & i g^{n-2}
\, (2 \pi)^4 \del^{(4)} \left( \sum_{i=1}^{n} p_i \right) \,
\widehat{A}_{MHV}^{(r_{-} s_{-})} (u)
\label{5-9}
\\
\widehat{A}_{MHV}^{(r_{-} s_{-})} (u) \! &=&
\!\! \sum_{\si \in \S_{n-1}}
\Tr (t^{c_1} t^{c_{\si_2}} t^{c_{\si_3}} \cdots t^{c_{\si_n}}) \,
\frac{ (u_r u_s )^4}{ (u_1 u_{\si_2})(u_{\si_2} u_{\si_3})
\cdots (u_{\si_n} u_1)}
\label{5-10}
\eeqar
where $u_i$ denotes the spinor momentum of the $i$-th gluon
($i=1,2, \cdots, n$). The elements $r$ and $s$ denote
the numbering indices of the negative-helicity gluons.
General amplitudes, the so-called non-MHV amplitudes, can be expressed in terms of
the MHV amplitudes $\widehat{A}_{MHV}^{(r_{-} s_{-})} (u)$.
Prescription for these expressions is called the Cachazo-Svrcek-Witten (CSW) rules.
For the next-to-MHV (NMHV) amplitudes, which contain three negative-helicity
gluons, the CSW rules can be expressed as
\beq
\widehat{A}^{(r_- s_- t_-)}_{NMHV} (u) = \sum_{(i,j)}
\widehat{A}^{(i_+ \cdots r_- \cdots s_- \cdots j_+ k_+)}_{MHV} (u)
\, \frac{\del_{kl}}{q_{ij}^2} \,
\widehat{A}^{(l_- \, (j+1)_+ \cdots t_- \cdots (i-1)_+)}_{MHV} (u)
\label{5-11}
\eeq
where the sum is taken over all possible choices for $(i, j)$ that
satisfy the ordering $i < r < s < j < t$. The momentum transfer
$q_{ij}$ between the two MHV vertices is given by
\beq
q_{ij} = p_i + p_{i+1 } + \cdots + p_{r} + \cdots + p_{s} + \cdots + p_{j}
\label{5-12}
\eeq
where $p$'s denote four-momenta of gluons as before.
General non-MHV amplitudes are then obtained by an iterative use of
the relation (\ref{5-11}).

We now notice a structural similarity between (\ref{2-27}) and (\ref{5-10}).
In terms of the supersymmetric holonomy operator in (\ref{5-1}),
an S-matrix functional $\F$ of gluon amplitudes can then be expressed as follows.
\beqar
\F \left[ a^{(h)c} \right] &= &  \widehat{W}^{(A)} \, \F_{MHV} \left[ a^{(h)c} \right]
\label{5-13}\\
\widehat{W}^{(A)} &=& \exp \left[
\int d^4 x d^4 y ~ \frac{\del_{kl}}{q^2} ~
\frac{\del}{\del a_{k}^{(+)}(x)} \otimes
\frac{\del}{\del a_{l}^{(-)}(y)}
  \right]
\label{5-14} \\
\F_{MHV} \left[ a^{(h)c} \right]
& = & \exp \left[ \frac{i}{g^2} \int d^4 x d^8 \th
~ \Theta_{R, \ga}^{(A)} (u; x ,\th) \right]
\label{5-15}
\eeqar
where $a^{(h)c}$ refers to a generic expression for $a_{i}^{(h_i)c_i}$
$(i = 1,2, \cdots )$, with $a_{i}^{(h_i)}$ being
$a_{i}^{(h_i)} =  t^{c_i} \, a_{i}^{(h_i)c_i}$ as in (\ref{2-26}).
An $x$-space representation of the operator, $a_{i}^{(h_i)} (x)$, is defined as
\beq
a_{i}^{(h_i)} (x) =
\int d \mu (p_i)
\,  a_{i}^{(h_i)} \,  e^{i x_\mu p_{i}^{\mu} }
\label{5-16}
\eeq
where $d \mu (p_i)$ denotes the Nair measure (\ref{5-8}).
An explicit expression for general gluon amplitudes
$\widehat{A}^{(1_{h_1} 2_{h_2} \cdots n_{h_n})} (u)$ is written as
\beqar
&&
\nonumber
\!\!\!\!\!\!\!
\left. \frac{\del}{\del a_{1}^{(h_1) c_1} (x_1) } \otimes
\frac{\del}{\del a_{2}^{(h_2) c_2} (x_2)} \otimes
\cdots \otimes \frac{\del}{\del a_{n}^{(h_n) c_n} (x_n)}
~ \F \left[  a^{(h)c}  \right] \right|_{a^{(h)c} (x)=0} \\
&=&
i g^{n-2} \widehat{A}^{(1_{h_1} 2_{h_2} \cdots n_{h_n})} (u)
\label{5-17}
\eeqar
where a set of $h_i = \pm$ $(i = 1, 2, \cdots , n)$ gives an arbitrary
helicity configuration.
The condition $a^{(h)} (x) = 0$ means that the remaining operators (or
source functions) should be evaluated as zero in the end of the calculation.

There are few remarks in the above expressions.
First of all, we choose the following normalization of the spinor momenta.
\beq
\oint_\ga d(u_1 u_2) \wedge d(u_2 u_3) \wedge \cdots \wedge d (u_m u_1) = 2^{m+1}
\label{5-18}
\eeq
Under a permutation of the numbering indices, a sign factor arises in
the above expression. We omit this sign factor as well as the factor
$(-1)^{h_1 + h_2 + \cdots + h_n}$ in (\ref{2-27})
since physical quantities are given by the squares of the amplitudes.

Secondly, we notice that
the Grassmann integral over $\th$'s picks up only the MHV amplitudes or vortices
since the integral vanishes unless we have the following factor
\beq
\left. \int d^8 \th  \, \xi_{r}^{1}\xi_{r}^{2}\xi_{r}^{3}\xi_{r}^{4}
\, \xi_{s}^{1}\xi_{s}^{2}\xi_{s}^{3}\xi_{s}^{4}
\right|_{\xi_{i}^{\al} = \th_{A}^{\al} u_{i}^{A} }
= \, (u_r u_s )^4
\label{5-19}
\eeq
We therefore find that the supersymmetric holonomy operator $\Theta_{R, \ga}(u; x, \th)$
naturally describes an S-matrix functional $\F_{MHV}$ for the MHV gluon amplitudes;
an explicit form of $\F_{MHV}$ is shown in (\ref{5-15}).
A Wick-like contraction operator $\widehat{W}^{(A)}$ in (\ref{5-14}) is introduced
so that we can obtain non-MHV amplitudes in terms of the MHV ones, following the
CSW prescription, in a language of functional derivatives.
This field theoretic description is convenient. For example,
in the expression (\ref{5-17}), the sum over $(i,j)$ in (\ref{5-11}) is
guaranteed by the functional derivatives acting on $\F$ and
the relation (\ref{5-19}).
This explains why the momentum transfer is denoted by
$q$ without the $(i, j)$ indices in (\ref{5-14}).
The relation (\ref{5-19}) also suggests that gluon amplitudes
vanish unless the helicity configuration can be factorized by
the MHV helicity configurations. Thus the helicity index is given by
$h_i = (+ , -)$, rather than the supersymmetric version
$\hat{h}_i = (0, \pm \frac{1}{2} , \pm )$.

Lastly, in obtaining the expression (\ref{5-17}), we also use the following relations.
\beqar
\nonumber
a_{1}^{(\pm)} \otimes a_{2}^{(h_2)} \otimes \cdots \otimes a_{m}^{(h_m)} \otimes a_{1}^{(0)}
& \equiv &
\hf [a_{1}^{(0)} , a_{1}^{(\pm)}] \otimes a_{2}^{(h_2)} \otimes \cdots \otimes a_{m}^{(h_m)} \\
&=&
\pm \hf a_{1}^{(\pm)} \otimes a_{2}^{(h_2)} \otimes \cdots \otimes a_{m}^{(h_m)}
\label{5-20}
\eeqar
This relation also holds under a permutation of the numbering indices.
Notice that the operators $a_{i}^{(\pm)}$ are, by construction, coupled with
the logarithmic one form $\om_{ij}$. Thus the indices $(1,2, \cdots , m)$ have
an antisymmetric property which we implicitly use in (\ref{5-20}).
This relation (\ref{5-20}) has also been used
in obtaining the expression (\ref{2-27}).

\noindent
\underline{Supersymmetrization of $\Theta_{R, \ga}^{(H)} (u, \bu)$}

In the following, we consider applications of the above expressions
to a gravitational theory.
In analogy with (\ref{5-1}), supersymmetrization of
$\Theta_{R, \ga}^{(H)} (u, \bu)$ in (\ref{4-1}) can be expressed as
\beq
\Theta_{R, \ga}^{(H)} (u , \bu; x, \th) = \Tr_{R, \ga} \, \Path \exp \left[
\sum_{m \ge 5}^{\infty} \oint_{\ga} \underbrace{H \wedge H \wedge \cdots \wedge H}_{m}
\right]
\label{5-21}
\eeq
where, as in (\ref{4-7})-(\ref{4-12}), $H$ is given by
\beqar
H &=&  \sqrt{8 \pi G_N} \sum_{1 \le i < j \le n} H_{ij} \, \om_{ij}
\label{5-22}\\
H_{ij} &=& \sum_{ \si \in \S_{r-1}} \sum_{ \tau \in \S_{n-r-2}}
\left(
\sum_{\hat{h}_{i \mu_i}} g_{i}^{(\hat{h}_{i \mu_i})} (x, \th) \otimes g_{j}^{(00)}
\right) \, \om_{\la_i \la_j}
\label{5-23} \\
\om_{ij} & = & d \log(u_i u_j) = \frac{d(u_i u_j)}{(u_i u_j)}
\label{5-24}
\eeqar
where $\hat{h}_{i \mu_i}$ denotes supersymmetrization of
$h_{i \mu_i} \equiv h_i h_{\mu_i}$, {\it i.e.},
$\hat{h}_{i \mu_i} \equiv \hat{h}_{i} \hat{h}_{\mu_i}$.
Either $\hat{h}_i$ or $\hat{h}_{\mu_i}$ represents a
helicity of a frame field with $\N = 4$ supersymmetry.
Thus the operator $g_{i}^{(\hat{h}_{i \mu_i})} (x, \th)$
consists of $\N = 8$ supermultiplits.
This corresponds to the fact that the gravitons are essentially given by
two copies of a frame field which we regard
as an analog of a gauge field in
$\N = 4$ super Yang-Mills theory.
Since gluons are expressed by the supertwistor variables,
gravitons can be described by the variables on $\cp^{3|4} \times \cp^{3|4}$.
(This space is not super ambitwistor space, which
is given by a product of $\cp^{3|4}$ and its dual with
a certain constraint, since here the two $\cp^{3|4}$'s
are the same in nature, respectively corresponding to two
frame fields by which a graviton is made of.)
The Grassmann variables $\th$ are now expressed as
$\th^{\al}_{A}$ with $A = 1, 2$ and $\al = 1,2, \cdots , 8 = \N$.
As in (\ref{5-5}), projection of these variables can be written as
\beq
\xi^\al \, = \, \th_{A}^{\al} u^A \, ~~~ (\al = 1,2, \cdots , 8)
\label{5-25}
\eeq
Since the $\N = 8$ multiplets are obtained from those of
$\N =4$, we can split the index $\al$ as
\beqar
\nonumber
&& \al \, = \, ( \al_1 , \al_2 ) \\
&& \al_1 = 1,2,3,4, ~~~
\al_2 = 5,6,7,8
\label{5-26}
\eeqar

The operator $g_{i}^{(\hat{h}_{i \mu_i})} (x, \th)$  in (\ref{5-23})
is then defined as
\beqar
g_{i}^{(\hat{h}_{i \mu_i})} (x, \th)
& = &
\left. \int d\mu (p_i) ~ g_{i}^{(\hat{h}_{i \mu_i})} ( \xi_{i} )
~  e^{ i x_\mu p_{i}^{\mu} }
\right|_{\xi_{i}^{\al} = \th_{A}^{\al} u_{i}^{A} }
\label{5-27}
\\
g_{i}^{(\hat{h}_{i \mu_i})} ( \xi_{i} ) &=&
T^{\mu_i} \, g_{i \mu_i}^{(\hat{h}_{i \mu_i})} ( \xi_{i} )
\, = \,
T^{\mu_i} \, e_{i}^{(\hat{h}_{i}) a} ( \xi_i ) \,
e_{\mu_i}^{(\hat{h}_{\mu_i}) a} ( \xi_i )
\label{5-28}
\eeqar
where $e_{i}^{(\hat{h}_{i}) a} (\xi_i)$'s are defined as
\beqar
\nonumber
e_{i}^{(+)a} (\xi_i) &=& e_{i}^{(+) a} \\ \nonumber
e_{i}^{\left( + \hf \right) a} (\xi_i) &=& \xi_{i}^{\al_1}
\, e_{i \, \al_1}^{ \left( + \hf \right) a} \\
e_{i}^{(0)a} (\xi_i) &=& \hf \xi_{i}^{\al_1} \xi_{i}^{\bt_1} \, e_{i \, \al_1 \bt_1}^{(0)a}
\label{5-29}
\\ \nonumber
e_{i}^{\left(- \hf \right)a} (\xi_i) &=& \frac{1}{3!} \xi_{i}^{\al_1}\xi_{i}^{\bt_1}
\xi_{i}^{\ga_1}
\ep_{\al_1 \bt_1 \ga_1 \del_1} \, {e_{i}^{ \del_1}}^{ \left( - \hf \right) a}
\\ \nonumber
e_{i}^{(-)a} (\xi_i) &=& \xi_{i}^{1} \xi_{i}^{2} \xi_{i}^{3} \xi_{i}^{4} \, e_{i}^{(-)a}
\eeqar
Similarly, $e_{\mu_i}^{(\hat{h}_{\mu_i}) a} (\xi_i)$'s are defined as
\beqar
\nonumber
e_{\mu_i}^{(+)a} (\xi_i) &=& e_{\mu_i}^{(+) a} \\ \nonumber
e_{\mu_i}^{\left( + \hf \right) a} (\xi_i) &=& \xi_{i}^{\al_2}
\, e_{\mu_i \, \al_2}^{ \left( + \hf \right) a} \\
e_{i}^{(0)a} (\xi_i) &=& \hf \xi_{i}^{\al_2} \xi_{i}^{\bt_2} \,
e_{\mu_i \, \al_2 \bt_2}^{(0)a}
\label{5-30}
\\ \nonumber
e_{\mu_i}^{\left(- \hf \right)a} (\xi_i) &=& \frac{1}{3!} \xi_{i}^{\al_2}\xi_{i}^{\bt_2}
\xi_{i}^{\ga_2}
\ep_{\al_2 \bt_2 \ga_2 \del_2} \, {e_{\mu_i}^{ \del_2}}^{ \left( - \hf \right) a}
\\ \nonumber
e_{\mu_i}^{(-)a} (\xi_i) &=& \xi_{i}^{5} \xi_{i}^{6} \xi_{i}^{7} \xi_{i}^{8} \,
e_{\mu_i}^{(-)a}
\eeqar
Notice that $\xi_i$ is common in the above expressions.
Namely, there appears no $\xi_{\mu_i}$.
This comes from the fact the graviton operator (\ref{5-27})
is a point-like operator in $\N = 8$ chiral superspace.
Alternatively, we can interpret $\xi_i$ as chiral superpartners of
the tangent-space coordinate $x_a$ ($a = 0,1,2,3$), with
spacetime not being supersymmetrized.
The latter interpretation can also be applied to Yang-Mills theories.

\noindent
\underline{An S-matrix functional for graviton amplitudes}

In the Yang-Mills case, general gluon amplitudes
are expressed as (\ref{5-17}), generated from the
S-matrix functional in (\ref{5-13}).
In the following, we obtain an analogous expression
for graviton amplitudes.
Gluon amplitudes are represented by
$\widehat{A}^{(1_{h_1} 2_{h_2} \cdots n_{h_n})} (u)$
in (\ref{5-17}).
In a momentum-space representation, the amplitudes should be
expressed as
\beqar
\A^{(1_{h_1} 2_{h_2} \cdots n_{h_n})} (u, \bu)
& = & i g^{n-2}
\, (2 \pi)^4 \del^{(4)} \left( \sum_{i=1}^{n} p_i \right) \,
\widehat{A}^{(1_{h_1} 2_{h_2} \cdots n_{h_n})} (u)
\label{5-31}
\\
\widehat{A}^{(1_{h_1} 2_{h_2} \cdots n_{h_n})} (u) &=&
\sum_{\si \in \S_{n-1}}
\Tr (t^{c_1} t^{c_{\si_2}} t^{c_{\si_3}} \cdots t^{c_{\si_n}}) ~
\C (1 \si_2 \si_3 \cdots \si_n)
\label{5-32}
\eeqar
where $\C (1 \si_2 \si_3 \cdots \si_n)$ are functions of
the Lorentz-invariant scalar products $(u_i u_j)$.
For the MHV amplitudes, an explicit form of these
can be written as (\ref{5-10}).
By use of the CSW rules, we can in principle obtain
$\C$'s of any helicity configurations.
In terms of such $\C$'s, we can
express tree-level graviton amplitudes as \cite{Bern:1998sv}
\beqar
\nonumber
\M^{(1_{h_{1 \mu_1}} 2_{h_{2 \mu_2}} \cdots n_{h_{n \mu_n}})} (u, \bu)
& = &  i ( 8 \pi G_N )^{\frac{n}{2} - 1} (-1)^{n+1}
\, (2 \pi)^4 \del^{(4)} \left( \sum_{i=1}^{n} p_i \right) \,
\\
&&
~ \times \,
\widehat{M}^{(1_{h_{1 \mu_1}} 2_{h_{2 \mu_2}} \cdots n_{h_{n \mu_n}})}
(u, \bu)
\label{5-33}
\\
\nonumber
\widehat{M}^{(1_{h_{1 \mu_1}} 2_{h_{2 \mu_2}} \cdots n_{h_{n \mu_n}})}
(u, \bu)
& = &  \sum_{\si \in \S_{r-1}} \sum_{\tau \in \S_{n-r-2}}
f(\si) \tilde{f} (\tau) ~ \C (1 2 \cdots n)
\\
\nonumber
&& \, \times \, \C(\si_2 \si_3 \cdots \si_{r} \, 1 \, n-1 \,
\tau_{r+1} \tau_{r+2} \cdots \tau_{n-2} \, n)
\\
&& \, + \, \P (23 \cdots n-2)
\label{5-34}
\eeqar
where $f(\si)$ and $\tilde{f} (\tau)$ are given by (\ref{4-21}),
with $m$ replaced by $n$.

In analogy with (\ref{5-15}),
an S-matrix functional for the MHV graviton amplitudes can be defined as
\beq
\F_{MHV} \left[ g_{i \mu_i}^{(h_{i \mu_i})} \right]
\, = \, \exp \left[ \frac{i}{8 \pi G_N} \int d^4 x \, d^{16} \th
~ \Theta_{R, \ga}^{(H)} (u, \bu ; x ,\th) \right]
\label{5-35}
\eeq
where $g_{i \mu_i}^{(h_{i \mu_i})}$ $(i = 1,2, \cdots )$ denotes an operator
or a source function associated with the expression
$g_{i}^{(h_{i \mu_i})} = T^{\mu_i}  g_{i \mu_i}^{( h_{i \mu_i} )}$.
As in (\ref{5-16}),
an $x$-space representation of the operator, $g_{i \mu_i}^{(h_{i \mu_i})} (x)$,
is defined as
\beq
g_{i \mu_i}^{(h_{i \mu_i})} (x) \, = \,
\int d \mu (p_i)
\,  g_{i \mu_i}^{(h_{i \mu_i})} \,  e^{i x \cdot p_{i} }
\label{5-36}
\eeq
where $d \mu (p_i)$ is the Nair measure (\ref{5-8}).

Labeling the two negative-helicity gravitons by $(s_{--}  t_{--} )$,
the MHV graviton amplitudes $\widehat{M}_{MHV}^{(s_{--} t_{--} )} (u, \bu)$
can be generated by (\ref{5-35}) as follows.
\beqar
\nonumber
&&
\!\!\!\!\!\!
\frac{\del}{\del g_{1 \mu_1}^{(++)}( x )} \otimes
\cdots \otimes \frac{\del}{\del g_{s \mu_s}^{(--) }( x )} \otimes
\cdots
\\
\nonumber
&&
~~~~~~~~~~~~
\left.
\cdots \otimes \frac{\del}{\del g_{t \mu_t}^{(--) }( x )} \otimes
\cdots \otimes \frac{\del}{\del g_{n \mu_n}^{(++)}( x )}
~ \F_{MHV} \left[ g_{i \mu_i}^{(h_{i \mu_i})} \right]
 \right|_{g_{i \mu_i}^{(h_{i \mu_i})} ( x ) =0}
\\
& = &
i ( 8 \pi G_N )^{\frac{n}{2}-1} \, \widehat{M}_{MHV}^{(s_{--} t_{--})} (u , \bu)
\label{5-37}
\eeqar
where we use (\ref{4-20}), (\ref{5-18}) and the following Grassmann integral
\beq
\left.
\int d^{16} \th \, \prod_{\al = 1}^{8} \xi_{s}^{\al}
\, \prod_{\bt = 1}^{8} \xi_{t}^{\bt} \right|_{\xi_{i}^{\al} = \th_{A}^{\al} u_{i}^{A}}
 =  \, ( u_s u_t )^8
\label{5-38}
\eeq

By use of the CSW rules in $\C$'s, we can straightforwardly extend the
above expressions to non-MHV cases.
An S-matrix functional for general tree-level graviton amplitudes
is then defined as
\beqar
\F \left[ g_{i \mu_i}^{(h_{i \mu_i})} \right]
&= &  \widehat{W}^{(H)} \, \F_{MHV} \left[ g_{i \mu_i}^{(h_{i \mu_i})} \right]
\label{5-39}\\
\widehat{W}^{(H)} &=& \exp \left[
\int d^4 x d^4 y ~ \frac{\del_{kl}}{q^2} ~
\frac{\del}{\del g_{k \mu_k}^{(++)}(x)} \otimes
\frac{\del}{\del g_{l \mu_l}^{(--)}(y)}
  \right]
\label{5-40}
\eeqar
where the meaning of $q$ in (\ref{5-40}) is exactly the same as
that in (\ref{5-14}) except that gluon momenta are
now generically replaced by graviton momenta.
In terms of the S-matrix functional (\ref{5-39}),
graviton amplitudes can generally be expressed as
\beqar
&&
\nonumber
\!\!\!\!\!\!\!
\left. \frac{\del}{\del g_{1 \mu_1}^{(h_{1 \mu_1 } )} (x_1) } \otimes
\frac{\del}{\del g_{2 \mu_2}^{(h_{2 \mu_2}) } (x_2)} \otimes
\cdots \otimes \frac{\del}{\del g_{n \mu_n}^{(h_{n \mu_n}) } (x_n)}
~ \F \left[  g_{i \mu_i}^{(h_{i \mu_i})}  \right]
\right|_{g_{i \mu_i}^{(h_{i \mu_i})} (x)=0} \\
&=&
i ( 8 \pi G_N )^{\frac{n}{2}-1}
\widehat{M}^{(1_{h_{1 \mu_1}} 2_{h_{2 \mu_2}} \cdots n_{h_{n \mu_n}})} (u , \bu )
\label{5-41}
\eeqar
where $h_{i \mu_i}$ $(i = 1, 2, \cdots , n)$ can take any
helicities in $(++ , --)$.
The rest of the helicity configurations are ruled out
due to the Grassmann integral (\ref{5-38}).
Notice that the particular assignment for the
index $\al$ in (\ref{5-26}) is crucial to extract
the helicities of $(++,--)$.
Without such an assignment,
the sates with $(+-, -+)$ helicities would emerge.
As mentioned before, the operators $g_{i \mu_i}^{(+-)}$
and $g_{i \mu_i}^{(-+)}$ represent stable and neutral
particles without mass and spin, which can be regarded
as a candidate for the origin of dark matter.
Observational evidence of dark matter and dark energy strongly suggests
that there should be operators like
$g_{i \mu_i}^{(+-)}$ and $g_{i \mu_i}^{(-+)}$
to be incorporated in a full gravitational theory.
In the present formalism, this can be carried out by relaxing
the assignment (\ref{5-26}) and construct a theory as a
square of $\N=4$ theories to include terms of
$g_{i \mu_i}^{(+-)}$ and $g_{i \mu_i}^{(-+)}$;
this formulation is currently under study.

Few other remarks on the expression (\ref{5-41}) is in order below.
One may wonder why there is a tedious label
$\mu_i$ for each of the helicity index $h_{i \mu_i}$.
This label is nothing but a Chan-Paton index (\ref{4-8}), playing the same role
as $c_i$ in the Yang-Mills case.
Thus it can actually be removed as in the expression (\ref{5-17}).
Lastly, the remaining $(-1)^{n+1}$ factor in (\ref{5-33}) can easily be
obtained by revising the definition of $e_{i}^{(\pm)}$ in (\ref{3-6}),
from $e_{i}^{(\pm)} = e_{i}^{(\pm)a} (\sqrt{2} p_i)^a$ to
$e_{i}^{(\pm)} = e_{i}^{(\pm)a} (\sqrt{- 2} p_i)^a$, which is the more
consistent with the expression in (\ref{3-2}).
This revision is however immaterial since, as discussed elsewhere,
physical quantities are obtained by the square of the amplitudes.

\section{Concluding remarks}

In the present paper, we construct a four-dimensional theory of
gravity in terms of a holonomy operator in twistor space
which has been introduced in the accompanying paper \cite{Abe:hol01}.
A gravitational holonomy operator $\Theta_{R, \ga}^{(H)}$
is defined by (\ref{4-1}) along with explicit expressions for $H$
in (\ref{4-7})-(\ref{4-11}).
We show that, as in the Yang-Mills case, an S-matrix functional for
scattering amplitudes of gravitons is naturally described by
a supersymmetric extension of $\Theta_{R, \ga}^{(H)}$.

Use of a holonomy formalism means that we construct a gravitational
theory as a gauge theory.
A Chan-Paton factor of the holonomy operator is
given by a trace over Poincar\'{e} algebra and Iwahori-Hecke algebra.
Owing to manifest Lorentz invariance of the formalism,
the Poincar\'{e} algebra is realized by translational operators
in tangent spaces. A trace over Iwahori-Hecke algebra, or a
braid trace, is realized by a sum over permutations of the numbering
indices for gravitons.
What is significant in this paper is that we clarify that
the Chan-Paton factor of $\Theta_{R, \ga}^{(H)}$ is in one-to-one
correspondence with a certain combinatoric factor in graviton amplitudes.

In an analysis of the Chan-Paton factor, we find that it can be
characterized by three distinct loops (up to isotopy), due to
an $SL(2, {\bf C})$ symmetry which is relevant to the Lorentz
invariance of the logarithmic one-form $\om_{ij}$ in (\ref{5-24}).
The characterization can be made by a notion of
ascending or descending order in the numbering indices
assigned for each loop.
Thus the very distinctiveness of the loops
requires that the total number of gravitons
should be more than or equal to five.
This is reflected in the definition of $\Theta_{R, \ga}^{(H)}$
as a condition of $m \ge 5$.
Since graviton amplitudes exists for $m \ge 3$, we can in fact
relax the above constraint to define the holonomy operator as
\beq
\Theta_{R, \ga}^{(H)} (u, \bu) = \Tr_{R, \ga} \, \Path \exp \left[
\sum_{m \ge 3} \oint_{\ga} \underbrace{H \wedge H \wedge \cdots \wedge H}_{m}
\right]
\label{6-1}
\eeq
with the same $H$ as defined in (\ref{4-7})-(\ref{4-11}).
One can easily check that this leads to
correct graviton amplitudes for $n = 3$ and $4$.
It is then possible to factorize the holonomy operator by the following
term.
\beq
\Tr_{R, \ga} \oint_\ga H \wedge H \wedge H
\label{6-2}
\eeq
As in the case of Yang-Mills theory (\ref{2-23}),
$H$ satisfies the integrability condition.
\beq
DH \, = \, dH - H \wedge H \, = \, 0
\label{6-3}
\eeq
Thus the term in (\ref{6-2}) can be interpreted as a
gravitational Chern-Simons term, with (\ref{6-3}) serving as
an Einstein equation.
This explains why Chern-Simons theory arises in so many areas of
physics, including Yang-Mills theories, gravitational theories, and
integrable models in general.\footnote{
This also gives an answer to a question we put
in the first footnote of the present paper.}

There are basically two ways of having a dual picture
between Yang-Mills theory and a gravitational theory in
the holonomy formalism.
These can easily be seen by a simple dimensional analysis as follows.
One way is to introduce a coupling constant $g_{f}$
for the frame fields $e_{i}^{(\pm)}$ and to express
the Newton constant in terms of $g_{f}$, which leads to
the relation $( 8 \pi G_N )^{\frac{1}{4}} = g_{f}$.
Thus the mass dimension of $g_{f}$ is given by $- \hf$.
This is not consistent with the fact that we have regarded $e_{i}^{(\pm)}$
as analogs of Yang-Mills fields which have dimensionless coupling constants.
We need to interpret $e_{i}^{(\pm)}$ as five-dimensional Yang-Mills fields
in order to overcome the discrepancy.
This suggests that we need to have
a notion which is similar to dimensional transmutation on $e_{i}^{(\pm)}$.
Such a concept is necessary if we like to
have an interpretation of gravity as a square of Yang-Mills theories.
Taking account of Chan-Paton factors, an appropriate coupling constant
can be written as $g_{F} = g_{f} M_{Pl}$ whose mass dimension is
given by $\hf$. The square of this constant has mass dimension 1 and
can be identified with a cosmological constant.
Thus it may be more convenient to use $g_{F}$ than $g_{f}$ in applications
to cosmology, however, there is still a dimensional discrepancy and
we need to have a concept like dimensional transmutation as well.

The other way is a rather direct one. Namely, we follow
our interpretation of gravity as a gauge theory and consider
$\sqrt{8 \pi G_N}$ multiplied by the Chan-Paton factor $T^{\mu_i} \sim M_{Pl}^2$
as a coupling constant of the gauge theory.
Since the mass dimension of $\sqrt{8 \pi G_N}  M_{Pl}^2$ is 1,
we can not relate this quantity to the Yang-Mills coupling constant $g$
unless we have a hidden Chen-Paton factor, with mass dimension 1, coupled to $g$.
Assuming such a factor, we can have the following correspondence
\beq
\sqrt{8 \pi G_N}  M_{Pl}^2 \, \Leftrightarrow \,  g (\bt M_{Pl} )
\label{6-4}
\eeq
where $\bt$ is some numerical constant.
This relation shows
an explicit weak-weak duality between Yang-Mills and gravitational theories.
The right-hand side in (\ref{6-4}) suggests a modification of
Yang-Mills theory with a tangent-space contribution which is analogous
to the gravitational theory constructed in this paper.
Such a modification is natural since Yang-Mills theory is also
invariant under spacetime translations and diffeomorphism in
the holonomy formalism.
The modification also implies breakings of conformal symmetry and
holomorphicity or chiral symmetry.
It is expected that this kind of symmetry breaking happens
spontaneously, providing a potentially new mechanism
to explain the origin of mass.
We do not know anything about such an interesting
direction of research yet but it is probably related to
deep algebraic problems.

As discussed in (\ref{2-24}), the Yang-Mills coupling constant $g$
is related to the Knizhnik-Zamolodchikov parameter. For $SU (N)$
gauge groups, $g$ is given by $g = {1 \over {1 + N}}$.
In such cases, an exact correspondence in (\ref{6-4}) determines
the value of $\bt$ as
\beq
\bt = \frac{\sqrt{8 \pi \hbar c}}{g} = \sqrt{\frac{8 \pi}{\al_{YM}}}
= \sqrt{8 \pi} (N+1)
\label{6-5}
\eeq
where $\al_{YM} = g^2 / \hbar c$ is a fine structure constant
for Yang-Mills theory.

As discussed in the Yang-Mills case, the
S-matrix functional (\ref{5-39}) may be utilized to generate
loop amplitudes without further modifications.
The ultraviolet finiteness of the theory, apart from Chan-Paton factors,
is guaranteed since it is constructed as a product of
$\N =4$ theories which are ultraviolet finite.
The Chan-Paton factors of the theory are given by
the left-hand side in (\ref{6-4}) times the number of gravitons.
Thus this part of the theory obviously diverges when the
number of gravitons is infinite.
In a practical calculation, however, we deal with a situation where the
number of gravitons is finite so that this divergence is not physically relevant.

Lastly, we would like to remark that massless spin-zero
particles, represented by $g_{i}^{(+-)}$ and $g_{i}^{(-+)}$
in (\ref{4-10}), are naturally incorporated in our construction of
a gravitational theory.
This shows that the holonomy formalism provides an interesting framework
for a search of the origin of dark matter and dark energy.

\vskip .3in
\noindent
{\bf Acknowledgments} \vskip .06in\noindent
I am deeply indebted to Professor V.P. Nair for
suggestion of braid structures in graviton amplitudes during my doctoral apprenticeship.
I would also like to thank the Yukawa Institute for Theoretical Physics at Kyoto University.
Discussions during the YITP workshop YITP-W-08-04 on
``Development of Quantum Field Theory and String Theory'' were useful for the present work.


\end{document}